\documentclass[twocolumn]{aastex631}
\usepackage{amsmath}     
\usepackage{wasysym}           
\usepackage{graphicx}
\usepackage{amssymb}
\usepackage{epstopdf}
\usepackage{mathrsfs}
\usepackage{anyfontsize}
\usepackage{natbib}
\usepackage{makecell}
\usepackage{enumitem}
\usepackage{color}
\usepackage{lipsum}
\usepackage{diagbox}

\shorttitle{From coasting to energy-conserving}
\shortauthors{E.~R.~Coughlin}
\begin{document}
\title{From coasting to energy-conserving: new self-similar solutions to the interaction phase of strong explosions}
\author[0000-0003-3765-6401]{Eric R.~Coughlin}
\affiliation{Department of Physics, Syracuse University, Syracuse, NY 13210, USA}

\email{ecoughli@syr.edu}

\begin{abstract}
Astrophysical explosions that contain dense and ram-pressure-dominated ejecta evolve through an interaction phase, during which a forward shock (FS), contact discontinuity (CD), and reverse shock (RS) form and expand with time. We describe new self-similar solutions that apply to this phase and are most accurate in the limit that the ejecta density is large compared to the ambient density. These solutions predict that the FS, CD, and RS expand at different rates in time and not as single temporal power-laws, are valid for explosions driven by steady winds and homologously expanding ejecta, and exist when the ambient density profile is a power-law with power-law index shallower than $\sim 3$ (specifically when the FS does not accelerate). We find excellent agreement between the predictions of these solutions and hydrodynamical simulations, both for the temporal behavior of the discontinuities and for the variation of the fluid quantities. The self-similar solutions are applicable to a wide range of astrophysical phenomena and -- although the details are described in future work -- can be generalized to incorporate relativistic speeds with arbitrary Lorentz factors. We suggest that these solutions accurately interpolate between the initial ``coasting'' phase of the explosion and the later, energy-conserving phase (or, if the ejecta is homologous and the density profile is sufficiently steep, the self-similar phase described in \citealt{chevalier82}). 
\end{abstract}


\section{Introduction}
\label{sec:intro}
The formation of a shockwave is an inevitable consequence of the injection of energy into a cold medium, making them ubiquitous in astrophysical settings. The dissipation of bulk kinetic energy at a shockwave occurs at the expense of the production of internal energy and radiation, making them not just ubiquitous but signposts of high-energy phenomena, including gamma-ray bursts (e.g., \citealt{rees92, meszaros97, sari98}), core-collapse (e.g., \citealt{woosley02, khatami23}) and type-Ia (e.g., \citealt{chevalier88, hillebrandt00}) supernovae, colliding-wind binaries (e.g., \citealt{stevens92, abaroa23}), galactic (e.g., \citealt{holzer70, thompson24}), stellar (e.g., \citealt{castor75, weaver77}), and AGN outflows (e.g., \citealt{begelman84, begelman89}), and tidal disruption events (TDEs; e.g., \citealt{rees88, gezari21, alexander16}). 

While an explosion and the outward motion of gas must initially be powered by some other (i.e., non-kinetic) energy reservoir, the thermal content of the expanding gas generally declines adiabatically and more rapidly than the kinetic energy, such that the ``ejecta'' of an explosion becomes cold and kinetically dominated soon after it is initiated. If, however, there is a medium surrounding the object at the time of the explosion, then the interaction between the expanding ejecta and that medium provides an additional source of dissipation and corresponding emission. This type of interaction is thought to power the late-time rebrightenings of some supernovae (i.e., type-IIn; e.g., \citealt{chevalier94, sollerman20}), is the most widely accepted mechanism for generating the afterglow associated with GRBs (e.g., \citealt{kumar15}), and can generally supplement -- and conceivably dominate -- the overall energetics of highly energetic, cosmic explosions (e.g., \citealt{dong16, arcavi17, andrews18, sollerman22}). 

During this ``interaction phase,'' the collision between the ambient gas and the ejecta creates a forward shock (FS) with radius $R_{\rm s}$ that advances into the ambient medium, a reverse shock (RS) with radius $R_{\rm r}$ that propagates into the ejecta, and a contact discontinuity (CD) with radius $R_{\rm c}$ that separates the two shocked fluids (see Figure 1 of \citealt{khatami23} for an illustration). Once the ejecta density becomes comparable to the ambient density (equivalent, in order of magnitude, to the condition that the swept-up mass be comparable to the initial explosion mass; e.g., \citealt{ostriker88}), we expect the FS to be set by the (potentially time-dependent, if a steady wind drives the outflow; e.g., \citealt{weaver77}) explosion energy, at which point the blastwave enters the self-similar Sedov-Taylor phase (\citealt{sedov59, taylor50}). \citet{chevalier82} showed that, when the ejecta is homologously expanding, there are also self-similar solutions that describe the initial interaction phase, and these solutions have time-independent ratios of $R_{\rm r}/R_{\rm c}$ and $R_{\rm s}/R_{\rm c}$. They also require a constant and specific ratio of the ejecta density, $\rho_{\rm ej}$, and the ambient density, $\rho_{\rm a}$, and are only attained if the (initial) ejecta density falls off as a power-law steeper than $\propto r^{-5}$. The origin of the latter constraint is that it is only for these steep power-law profiles that the kinetic energy of the ejecta is infinite\footnote{A corollary of this constraint is that the cylindrical analogs of the \citet{chevalier82} solutions presumably exist for profiles steeper than $\propto s^{-4}$, where $s$ is cylindrical radius.}; flatter profiles have a finite energy and must be governed by the conservation of that energy at sufficiently late times. 

Here we show that there are distinct self-similar solutions that describe the interaction phase, which are valid when the density of the ejecta, $\rho_{\rm ej}$, exceeds the density of the ambient medium, $\rho_{\rm a}$, and are most accurate when $\rho_{\rm ej} \gg \rho_{\rm a}$. During this phase the forward shock ``coasts'' at a fixed value of the ejecta velocity, meaning that the self-similar FS shell thickness, $\Delta_{\rm s} = R_{\rm s}/R_{\rm c}-1$, is constant in time. The thickness of the reverse shock, $\Delta_{\rm r} = 1-R_{\rm r}/R_{\rm c}$, scales as $\sqrt{\rho_{\rm a}/\rho_{\rm ej}}$ and is therefore generally time-dependent, implying that the self-similar solutions possess different rates of expansion of the FS, CD, and RS. Our solutions are valid while $\Delta_{\rm r} \lesssim \Delta_{\rm s}$, after which we expect the solution to transition either to \citet{chevalier82}'s if the ejecta density profile is sufficiently steep, or to the Sedov-Taylor solution when it is less so (or that of \citealt{weaver77} for wind-powered explosions), and therefore interpolate between the coasting and energy-conserving phases of a strong explosion. 

In Section \ref{sec:basic} we provide motivation and physical arguments for the existence and properties of the self-similar solutions, and in Section \ref{sec:solutions} we derive the solutions and provide comparisons to numerical simulations. We discuss various aspects of the solutions and analyze applications in Section \ref{sec:discussion} before briefly summarizing in Section \ref{sec:summary}.

\section{Basic considerations}
\label{sec:basic}
Let a spherically symmetric and kinetic-energy-dominated outflow (the ``ejecta'') impinge upon a static and cold ambient medium. At the moment that the impact occurs, we let the ejecta have density $\rho_{\rm ej}$ and velocity $V_{\rm ej}$ at the radius of impact, and the ambient medium have density $\rho_{\rm a}$ (also at that radius). Then the initial velocities of the FS, RS, and CD from the jump conditions are, respectively, 
\begin{equation}
\begin{split}
&V_{\rm s} = \frac{\gamma+1}{2}\frac{1}{1+\sqrt{\rho_{\rm a}/\rho_{\rm ej}}}V_{\rm ej}, \\
&V_{\rm r} = \frac{1-\frac{\gamma-1}{2}\sqrt{\rho_{\rm a}/\rho_{\rm ej}}}{1+\sqrt{\rho_{\rm a}/\rho_{\rm ej}}}V_{\rm ej}, \\
&V_{\rm c} = \frac{1}{1+\sqrt{\rho_{\rm a}/\rho_{\rm ej}}}V_{\rm ej}. \label{Vinits}
\end{split}
\end{equation}
Here $\gamma$ is the adiabatic index of the gas (we are assuming that both the shocked ejecta and ambient fluids are characterized by the same adiabatic index, though this assumption could be trivially relaxed). 

Following the initial interaction, the RS, CD, and FS spatially separate from one another owing to their different velocities, and the shocked fluid variables at the RS and the FS will depend on the ejecta and ambient properties at those respective radii, i.e., the spatial variation of the ejecta and ambient fluids is relevant in addition to their values characterizing the initial impact. The fluid variables at the FS and RS that arise from the jump conditions are then:
\begin{equation}
\begin{split}
&v(R_{\rm s}) = \frac{2}{\gamma+1}V_{\rm s}, \,\,\, v(R_{\rm r}) = \frac{2}{\gamma+1}V_{\rm r}+\frac{\gamma-1}{\gamma+1}v_{\rm ej}(R_{\rm r}), \\
&\rho(R_{\rm s}) = \frac{\gamma+1}{\gamma-1}\rho_{\rm a}(R_{\rm s}), \,\,\, \rho(R_{\rm r}) = \frac{\gamma+1}{\gamma-1}\rho_{\rm ej}(R_{\rm r})\\
&p(R_{\rm s}) = \frac{2}{\gamma+1}\rho_{\rm a}(R_{\rm s})V_{\rm s}^2, \\
&\quad\quad \quad \quad \quad  p(R_{\rm r}) = \frac{2}{\gamma+1}\rho_{\rm ej}(R_{\rm r})\left(v_{\rm ej}(R_{\rm r})-V_{\rm r}\right)^2. \label{jumpbcs}
\end{split}
\end{equation}
Here $v$ is the fluid velocity and $p$ is the pressure, and we have incorporated the spatial dependence on the ejecta and ambient densities by explicitly writing the radii at which they are evaluated. We have also written the ejecta velocity as $v_{\rm ej}$, i.e., $v_{\rm ej}(r,t)$ refers to the spatially and temporally varying velocity profile of the inner ejecta ($V_{\rm ej}$, being the velocity that the leading edge of the ejecta {would} have if the ratio $\rho_{\rm a}/\rho_{\rm ej} \equiv 0$, has distinct physical significance and will be relevant, and we therefore do not re-use this variable). 

While Equation \eqref{Vinits} is only strictly valid for the initial interaction between cold ejecta impinging on cold ambient gas, it shows that the RS-CD-FS structure expands (or, if the ambient density profile is sufficiently steep, contracts) as a function of the ratio $\rho_{\rm a}/\rho_{\rm ej}$. If $R_{\rm s}$, $R_{\rm c}$, and $R_{\rm r}$ vary in such a way that $\rho_{\rm a}(R_{\rm s})/\rho_{\rm ej}(R_{\rm r})$ is time-independent, \citet{chevalier82} showed that self-similar solutions can be found that satisfy $R_{\rm r}/R_{\rm c} \propto R_{\rm s}/R_{\rm c}\sim$ const.~and $R_{\rm c}\propto t^{\alpha}$ with $\alpha$ a constant. But in general -- and for ejecta density profiles that do not initially conform to steep power-laws in radius -- such solutions cannot be found, and the RS, CD, and FS propagate at different speeds. 

There is, however, another scenario in which the interaction between the ejecta and ambient gas simplifies and for which approximate self-similar solutions could plausibly exist: if $\rho_{\rm a}/\rho_{\rm ej} \ll 1$, Equation \eqref{Vinits} shows $V_{\rm r} = V_{\rm ej}\left(1+\mathcal{O}\left[\sqrt{\rho_{\rm a}/\rho_{\rm ej}}\right]\right)$ and $V_{\rm c} = V_{\rm ej}\left(1+\mathcal{O}\left[\sqrt{\rho_{\rm a}/\rho_{\rm ej}}\right]\right)$, such that as $\rho_{\rm a}/\rho_{\rm ej} \rightarrow 0$, the RS ``piles up'' into the CD and the velocity is continuous to the CD; this behavior is reasonable, because when $\rho_{\rm ej}/\rho_{\rm a} \gg 1$, we would expect the ejecta to expand uninhibited by the ambient medium. In this same limit, the velocity of the ejecta is unaltered by the interaction, meaning that the velocity at the RS satisfies $v_{\rm ej}(R_{\rm r}) = V_{\rm ej}$ and is given by its initial value, i.e., $V_{\rm r} = V_{\rm c} = V_{\rm ej}$ for all $t > 0$. In this ``coasting'' regime (e.g., \citealt{ostriker88}) during which the CD has a constant speed, the FS must have $V_{\rm s} \propto V_{\rm ej} =$ const.~to not be overtaken by the CD (i.e., it cannot decelerate, and we are assuming it does not accelerate; see Section \ref{sec:discussion} for a description of where the latter breaks down) and $\Delta_{\rm s} \equiv R_{\rm s}/R_{\rm c}-1 =$ const. Self-similar solutions would then yield the values of $\Delta_{\rm s}$ and $V_{\rm s}/V_{\rm ej}$. 

Additionally, the difference between the velocity of the CD, $V_{\rm c}$, and the velocity at the leading edge of the ejecta in the limit that $\rho_{\rm a}/\rho_{\rm ej} \equiv 0$, $V_{\rm ej}$, constitutes a smallness parameter in this regime, and from Equation \eqref{Vinits} scales as $\propto \sqrt{\rho_{\rm a}/\rho_{\rm ej}}\equiv \delta$. As the ejecta expands, this ratio can be evaluated at the CD by extrapolating the inner and outer density profiles, and $\delta$ will therefore depend on time. The same smallness parameter characterizes the difference between $V_{\rm r}$ and $V_{\rm c}$ in this limit, such that the relative shell thickness between the CD and the RS, $\Delta_{\rm r} = 1-R_{\rm r}/R_{\rm c}$, scales as $\delta$. To satisfy the jump conditions across the RS and be continuous across the CD, the pressure within the RS shell must then satisfy $p \propto \rho_{\rm ej}V_{\rm c}^2\Delta_{\rm r}^2$. Because of the time dependence contained in the shell thickness, exact self-similar solutions for the RS during this phase likely do not exist\footnote{There is one exception to this, which arises when the ejecta is powered by a steady wind and when the ambient medium is wind-fed, and in which case the smallness parameter $\delta$ is independent of time; see Section \ref{sec:exact} below.}, but these observations suggest that approximate self-similar solutions could be found for which the fluid variables are written as power series in the small quantity $\delta$. 

We derive these self-similar solutions in the next section. The zeroth-order (in $\delta$) FS solutions (Section \ref{sec:forward}) are contained in the Sedov space and simply have zero acceleration, and depend on neither the density nor the velocity profile of the inner ejecta. Despite being fairly trivial, we have not seen these solutions discussed in this context or this generally\footnote{\citet{gruzinov03} and \citet{kushnir10} describe solutions with this property -- their ``type III'' similarity solutions -- when considering explosions in steep power-law media ($n \sim 3$), arguing that this is the correct limit to take for the Sedov problem and that the CD acts as a constant-velocity piston, interior to which the solution does not behave self-similarly. In our case, the solution interior to the RS (and CD when $\rho_{\rm a}/\rho_{\rm ej}\rightarrow 0$) is trivial and self-similar, and we expect our solutions to apply to all $n$ for which the FS decelerates ($n \lesssim 3$; see the discussion in Section \ref{sec:discussion}) and while $\Delta_{\rm r}\lesssim \Delta_{\rm s}$. We also suggest that our FS similarity solution is of the first type, because the deceleration rate of the FS -- zero -- is inferred on physical grounds.}. The RS solutions (Section \ref{sec:reverse}) are constrained simultaneously by the (known, $\propto \delta$) expansion rate of the shell $\Delta_{\rm r}$ and the continuity of the velocity and pressure at the CD. The latter are satisfied for special deceleration parameters that act as eigenvalues, but instead of permitting the smooth passage of the fluid variables through a sonic point (as is typically the case for type-II similarity solutions; \citealt{zeldovich67}), they satisfy boundary conditions at the CD. The solutions apply to both wind-driven and homologously-expanding-ejecta-driven explosions, both of which we analyze. Finally, we construct first-order ``corrections'' to the FS and shocked fluid (Section \ref{sec:corrections}), which describe the response of the FS to finite $\delta$; analogous to the RS, these require a precise growth rate for $\Delta_{\rm s}$ (alongside the constant and self-similar width) that serves to satisfy continuity conditions at the CD. 

Before constructing and analyzing the self-similar solutions, we establish and define the relevant radii and the physical setup that will be used throughout the remainder of the paper. Three time-dependent radii that were already introduced are the RS radius $R_{\rm r}(t)$, the CD radius $R_{\rm c}(t)$, and the FS radius $R_{\rm s}(t)$. From these we construct the relative FS shell thickness, $\Delta_{\rm s} = R_{\rm s}/R_{\rm c}-1$, and the relative RS shell thickness, $\Delta_{\rm r} = 1-R_{\rm r}/R_{\rm c}$, both of which are time-dependent. There is a fourth radius, $R_{\rm ej}(t)$, that is the radius of the leading edge of the ejecta in the limit that the ratio of the ambient to ejecta density is identically zero (which would coincide with the radius of the CD and the RS in the same limit), which we write as
\begin{equation}
R_{\rm ej}(t) = R_{\rm i}\left(1+\frac{V_{\rm ej}\,t}{R_{\rm i}}\right).
\end{equation}
The radius $R_{\rm i}$ is the radius of $R_{\rm ej}$ at $t = 0$ and $V_{\rm ej}$ is the (constant) velocity of the leading edge of the ejecta, which was also introduced above. $R_{\rm i}$ is distinguished from (and is not equal to) the radius of the CD at $t = 0$, because all three discontinuities and the ejecta radius are coincident at the origin at time $t = -R_{\rm i}/V_{\rm ej}$. The inner ejecta (i.e., interior to the RS) is cold and expanding ballistically, such that the velocity profile is conserved in a Lagrangian sense, i.e., each fluid element preserves its initial velocity. The two types of outflow we consider are homologously expanding ejecta and a constant-velocity wind, such that the Eulerian velocity at the RS is
\begin{equation}
v_{\rm ej}(R_{\rm r}) = \\
\begin{cases}
V_{\rm ej}\frac{R_{\rm r}(t)}{R_{\rm ej}(t)}, & \textrm{ homologous expansion }\\
V_{\rm ej}, & \textrm{ time-steady wind }
\end{cases}.
\end{equation}
In the case of a wind, time steadiness demands that the ejecta density fall off with Eulerian radius as $\propto r^{-2}$. The spatial dependence of homologously expanding ejecta can be more general, e.g., \citet{chevalier82} considered the case where the ejecta has a power-law density profile. However, in the limit that we are analyzing where the ejecta is nearly unaltered by the ambient gas, it is only the time dependence near the RS (which, in turn, is nearly equal to $R_{\rm c}$ and $R_{\rm ej}$) that enters, such that $\rho_{\rm ej} \propto r^{-3}$ to leading order in $\delta$. We therefore have
\begin{equation}
\begin{split}
\rho_{\rm ej}(R_{\rm r}) = &
\begin{cases}
\rho_{\rm ej}\left(\frac{R_{\rm r}(t)}{R_{\rm i}}\right)^{-3}, & \textrm{homologous expansion} \\ 
\rho_{\rm ej}\left(\frac{R_{\rm r}(t)}{R_{\rm i}}\right)^{-2}, & \textrm{time-steady wind}
\end{cases}
\\
&\equiv \rho_{\rm ej}\left(\frac{R_{\rm r}(t)}{R_{\rm i}}\right)^{-m}. \label{rhoej}
\end{split}
\end{equation}
Finally, we will let the ambient density profile conform to a power-law in radius, such that the ambient density evaluated at the FS is
\begin{equation}
\rho_{\rm a}(R_{\rm s}) = \rho_{\rm a}\left(\frac{R_{\rm s}(t)}{R_{\rm i}}\right)^{-n}. \label{rhoamb}
\end{equation}
In Equations \eqref{rhoej} and \eqref{rhoamb} we re-used $\rho_{\rm ej}$ and $\rho_{\rm a}$ to refer to the ejecta and ambient densities measured at $R_{\rm i}$ for simplicity of notation; in all of what follows, any appearance of $\rho_{\rm ej}$ and $\rho_{\rm a}$ will refer to these normalizations, and any explicit time dependence will be included.

\begin{figure*}[htbp] 
   \centering
   \includegraphics[width=\textwidth]{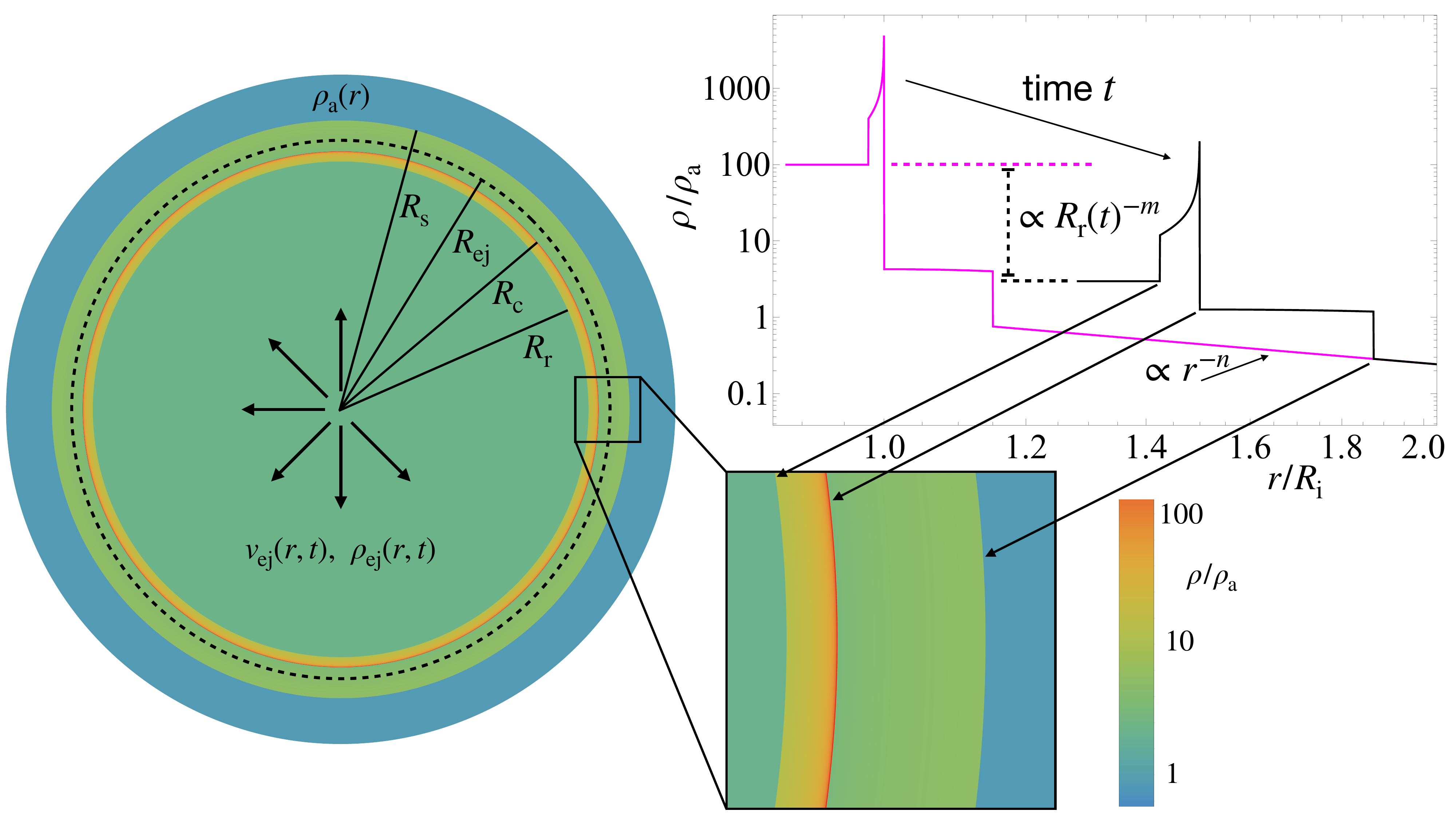} 
   \caption{A diagram of the relevant radii, being the reverse shock (RS), the contact discontinuity (CD), and the forward shock (FS), as well as the ejecta radius $R_{\rm ej}$; the latter is the radius that the CD and RS would have in the limit that the ratio of the ambient to ejecta density were identically zero. The top-right inset shows how the ejecta density declines with time as $R_{\rm r}^{-m}$, where $m = 2$ for a steady wind and $m = 3$ for homologously expanding ejecta, as well as the power-law decline of the ambient density as $\propto r^{-n}$, and the overall variation of the density throughout the shocked FS and RS shells. The inset on the bottom-right shows a zoom-in on the RS-CD-FS structure, as well as the locations of the three discontinuities in the top-right inset.}
   \label{fig:shock_schematic}
\end{figure*}

Figure \ref{fig:shock_schematic} illustrates the four radii in the problem as well as the various regions, including the unshocked ejecta, shocked ejecta, shocked ambient gas, and ambient gas. The inset on the top-right illustrates the variation of the density throughout the entire shell, as well as the temporal decline of the ejecta density with time (as $\propto R_{\rm r}^{-m}$), and the spatial decline of the ambient density (as $\propto r^{-n}$).

\section{Self-similar solutions for interacting explosions}
\label{sec:solutions}
\subsection{Forward shock}
\label{sec:forward}
We assume that the ratio of the ambient to ejecta density is sufficiently small that $V_{\rm c} \simeq V_{\rm ej}$. We define the fluid and self-similar variables as
\begin{equation}
\begin{split}
&v = V_{\rm c}f_{\rm s}(\eta), \\
&\rho = \rho_{\rm a}\left(\frac{R_{\rm c}}{R_{\rm i}}\right)^{-n}g_{\rm s}(\eta), \\
&p = \rho_{\rm a}V_{\rm c}^2\left(\frac{R_{\rm c}}{R_{\rm i}}\right)^{-n}h_{\rm s}(\eta), \\
&\eta = \frac{r/R_{\rm c}-1}{R_{\rm s}/R_{\rm c}-1} \equiv \frac{r/R_{\rm c}-1}{\Delta_{\rm s}}, 
\label{fluiddefs} 
\end{split}
\end{equation}
where $0 \le \eta \le 1$ and $\Delta_{\rm s}$ a constant. Then the fluid equations in spherical symmetry, 
\begin{equation}
\begin{split}
&\frac{\partial \rho}{\partial t}+\frac{1}{r^2}\frac{\partial}{\partial r}\left[r^2\rho v\right] = 0, \\
&\frac{\partial v}{\partial t}+v\frac{\partial v}{\partial r}+\frac{1}{\rho}\frac{\partial p}{\partial r} = 0, \\
&\frac{\partial}{\partial t}\ln\left(\frac{p}{\rho^{\gamma}}\right)+v\frac{\partial}{\partial r}\ln\left(\frac{p}{\rho^{\gamma}}\right) = 0,
\end{split}
\end{equation}
respectively become
\begin{equation}
\begin{split}
-ng_{\rm s}&-\frac{1}{\Delta_{\rm s}}\left(1+\Delta_{\rm s}\eta\right)\frac{dg_{\rm s}}{d\eta} \\ 
&+\frac{1}{\Delta_{\rm s}\left(1+\Delta_{\rm s}\eta\right)^2}\frac{d}{d \eta}\left[\left(1+\Delta_{\rm s}\eta\right)^2f_{\rm s}g_{\rm s}\right] = 0, \label{ssfsg}
\end{split}
\end{equation}
\begin{equation}
-\left(1+\Delta_{\rm s}\eta-f_{\rm s}\right)\frac{d f_{\rm s}}{d \eta}+\frac{1}{g_{\rm s}}\frac{d h_{\rm s}}{d \eta} = 0, \label{ssfsf}
\end{equation}
\begin{equation}
n\left(\gamma-1\right)-\frac{1}{\Delta_{\rm s}}\left(1+\Delta_{\rm s}\eta-f_{\rm s}\right)\frac{d}{d \eta}\ln\left(\frac{h_{\rm s}}{g_{\rm s}^{\gamma}}\right) = 0, \label{ssfsh}
\end{equation}
while the boundary conditions at the shock, from Equation \eqref{jumpbcs}, are
\begin{equation}
\begin{split}
&f_{\rm s}(1) = \frac{2}{\gamma+1}\left(1+\Delta_{\rm s}\right), \\
&g_{\rm s}(1) = \frac{\gamma+1}{\gamma-1}\left(1+\Delta_{\rm s}\right)^{-n}, \\
&h_{\rm s}(1) = \frac{2}{\gamma+1}\left(1+\Delta_{\rm s}\right)^{2-n}, 
\label{bcshock}
\end{split}
\end{equation}
and that at the CD is
\begin{equation}
f_{\rm s}(0) = 1. \label{bccd}
\end{equation}
Equations \eqref{ssfsg} -- \eqref{ssfsh} are solved by integrating from the shock ($\eta = 1$) and iterating on $\Delta_{\rm s}$ until Equation \eqref{bccd} is satisfied\footnote{Because $\Delta_{\rm s}$ is a constant, the self-similar variable $\eta$ is just linearly related to the usual Sedov variable, $\xi = r/R_{\rm s}$. Therefore, in practice it is more efficient to solve the usual Sedov-Taylor self-similar equations (e.g.~and using these definitions and variables, Equations 13 and 14 of \citealt{coughlin22}), from which the value of $\Delta_{\rm s}$ is delimited and obtained ``for free'' by where the self-similar solution terminates.}. 

\subsection{Reverse shock}
\label{sec:reverse}
From our considerations in Section \ref{sec:basic}, we expect the radius of the CD to vary as
\begin{equation}
R_{\rm c} = R_{\rm ej}(t)\left\{1-\kappa_{\rm c}\delta(R_{\rm c})\right\}, \label{Rckap}
\end{equation}
where $\kappa_{\rm c}$ is an unknown constant and
\begin{equation}
\delta(R_{\rm c}) = \sqrt{\frac{\rho_{\rm a}}{\rho_{\rm ej}}}\left(\frac{R_{\rm c}(t)}{R_{\rm i}}\right)^{\frac{m-n}{2}}.
\end{equation}
We define the RS radius $R_{\rm r}$ and self-similar variable as
\begin{equation}
\begin{split}
&R_{\rm r} = R_{\rm c}\left(1-\Delta_{\rm r}\right) \equiv R_{\rm c}\left\{1-\kappa_{\rm r}\delta(\tau)\right\}, \\
&\eta = \frac{r/R_{\rm c}-1}{1-R_{\rm r}/R_{\rm c}} = \frac{r/R_{\rm c}-1}{\Delta_{\rm r}}, \label{fluiddefsRr}
\end{split}
\end{equation}
where $\kappa_{\rm r}$ is another unknown parameter and\footnote{Note that the $\eta$ appearing in the definition of the reverse-shock fluid variables is not the same as that for the forward-shock fluid variables; we omitted subscripts from the various $\eta$'s for ease of notation.} $-1 \le \eta \le 0$. 
Because $\Delta_{\rm r}$ is time-dependent, the usual self-similar parameterization of the fluid variables (e.g., that used for the FS) will not lead to self-consistent equations for $f_{\rm r}$, $g_{\rm r}$, and $h_{\rm r}$. However, and again from the physical considerations in Section \ref{sec:basic}, we expect the following to hold to leading order in $\Delta_{\rm r}$:
\begin{equation}
\begin{split}
&v = V_{\rm c}\left\{1+\Delta_{\rm r}f_{\rm r}(\eta)\right\}, \\
&\rho = \rho_{\rm ej}\left(\frac{R_{\rm c}}{R_{\rm i}}\right)^{-m}g_{\rm r}(\eta) \\
&p = \rho_{\rm ej}\left(\frac{R_{\rm c}}{R_{\rm i}}\right)^{-m}V_{\rm c}^2\Delta_{\rm r}^2h_{\rm r}(\eta). \\
 \label{ssRSdefs}
\end{split}
\end{equation}
The self-similar equations are derived by keeping lowest-order terms in $\Delta_{\rm r}$ in the fluid equations and using the definitions of $R_{\rm c}$, $R_{\rm r}$, and $\Delta_{\rm r}$ in terms of $\delta$, and are 
\begin{equation}
\begin{split}
\left(2-m\right)g_{\rm r}-\left(1+\frac{m-n}{2}\right)\eta\frac{dg_{\rm r}}{d\eta}+\frac{d}{d\eta}\left[f_{\rm r}g_{\rm r}\right] = 0, \label{ss1}
\end{split}
\end{equation}
\begin{equation}
\begin{split}
&-\frac{\kappa_{\rm c}}{\kappa_{\rm r}}\left(\frac{m-n}{2}\right)\left(1+\frac{m-n}{2}\right) +\frac{m-n}{2}f_{\rm r} \\ 
&+\left(f_{\rm r}-\eta-\frac{m-n}{2}\eta\right)\frac{df_{\rm r}}{d\eta}+\frac{1}{g_{\rm r}}\frac{dh_{\rm r}}{d\eta} = 0, \label{ss2}
\end{split}
\end{equation}
\begin{equation}
m\gamma-n+\left(f_{\rm r}-\eta-\frac{m-n}{2}\eta\right)\frac{d}{d\eta}\ln\left(\frac{h_{\rm r}}{g_{\rm r}^{\gamma}}\right) = 0. \label{ss3}
\end{equation}
The boundary conditions at the shock can be determined from the jump conditions \eqref{jumpbcs} and are 
\begin{equation}
\begin{split}
f_{\rm r}(-1) = \frac{2}{\gamma+1}&\left(1+\frac{m-n}{2}\right)\left(\frac{\gamma-1}{2}\frac{\kappa_{\rm c}}{\kappa_{\rm r}}-1\right) \\ 
&-\frac{\gamma-1}{\gamma+1}\left(1+\frac{\kappa_{\rm c}}{\kappa_{\rm r}}\right)\left(m-2\right), \label{ssbc1}
\end{split}
\end{equation}
\begin{equation}
h_{\rm r}(-1) = \frac{2}{\gamma+1}\left(1+\frac{\kappa_{\rm c}}{\kappa_{\rm r}}\right)^2\left(1+\frac{m-n}{2}-\left(m-2\right)\right)^2, \label{ssbc2}
\end{equation}
\begin{equation}
g_{\rm r}(-1) = \frac{\gamma+1}{\gamma-1}. \label{ssbc3}
\end{equation}
The factors proportional to $\left(m-2\right)$ account for the additional terms that enter in the case of homologously expanding ejecta ($m = 3$) vs.~a steady wind ($m = 2$), i.e., Equations \eqref{ssbc1} -- \eqref{ssbc3} are correct for both $m = 2$ and $m = 3$. The continuity of the velocity at the CD also requires
\begin{equation}
f_{\rm r}(0) = 0. \label{ssbc4}
\end{equation}

Equations \eqref{ss1} -- \eqref{ssbc4} depend only on the unknown ratio $\kappa_{\rm c}/\kappa_{\rm r}$; as for the FS shell thickness $\Delta_{\rm s}$, $\kappa_{\rm c}/\kappa_{\rm r}$ acts as an eigenvalue that can be determined iteratively by shooting from the RS until the condition $f_{\rm r}(0) = 0$ is met. The individual values of $\kappa_{\rm r}$ and $\kappa_{\rm c}$ are recovered by requiring pressure continuity at the CD: since the pressure at the CD from the forward-shocked material is given by the self-similar parameterization in Section \ref{sec:forward}, we have
\begin{equation}
\begin{split}
&\rho_{\rm ej}R_{\rm c}^{-m}V_{\rm c}^2\kappa_{\rm r}^2\delta^2h_{\rm r}(0) = \rho_{\rm a}R_{\rm c}^{-n}V_{\rm c}^2h_{\rm s}(0) \\
&\Rightarrow \quad \kappa_{\rm r} = \sqrt{\frac{h_{\rm s}(0)}{h_{\rm r}(0)}}, \label{bcp}
\end{split}
\end{equation}
thus closing the system.

In imposing Equation \eqref{bcp} we assumed that $h_{\rm r}$ is finite at the CD. In Appendix \ref{sec:asymptotic} we analyze the asymptotic behavior of the solutions near the CD and show that the pressure does indeed converge to a finite value, but the convergence is slow -- especially as $n$ approaches 3 for $m = 3$.

\subsection{Corrections to the forward shock}
\label{sec:corrections}
The self-similar solutions for the FS in Section \ref{sec:forward} set $\rho_{\rm a}/\rho_{\rm ej} = 0$ and are independent of this parameter. However, the FS will be modified analogously to the CD (i.e., Equation \ref{Rckap}) owing to the finite value of $\rho_{\rm a}/\rho_{\rm ej}$, and there will be modifications to $\Delta_{\rm s}$ that scale as $\delta$. We account for these effects by letting $f_{\rm s}(\eta) \rightarrow f_{\rm s}(\eta)+\delta(\tau)f_1(\eta)$, and similarly for the functions $g_{\rm s}$ and $h_{\rm s}$, in Equation \eqref{fluiddefs}, and letting
\begin{equation}
\Delta_{\rm s}(\tau) \rightarrow \Delta_{\rm s}+\kappa_{\rm s}\delta(\tau),
\end{equation}
with $\kappa_{\rm s}$ an unknown parameter.

\begin{figure*}[t!] 
   \centering
   \includegraphics[width=0.495\textwidth]{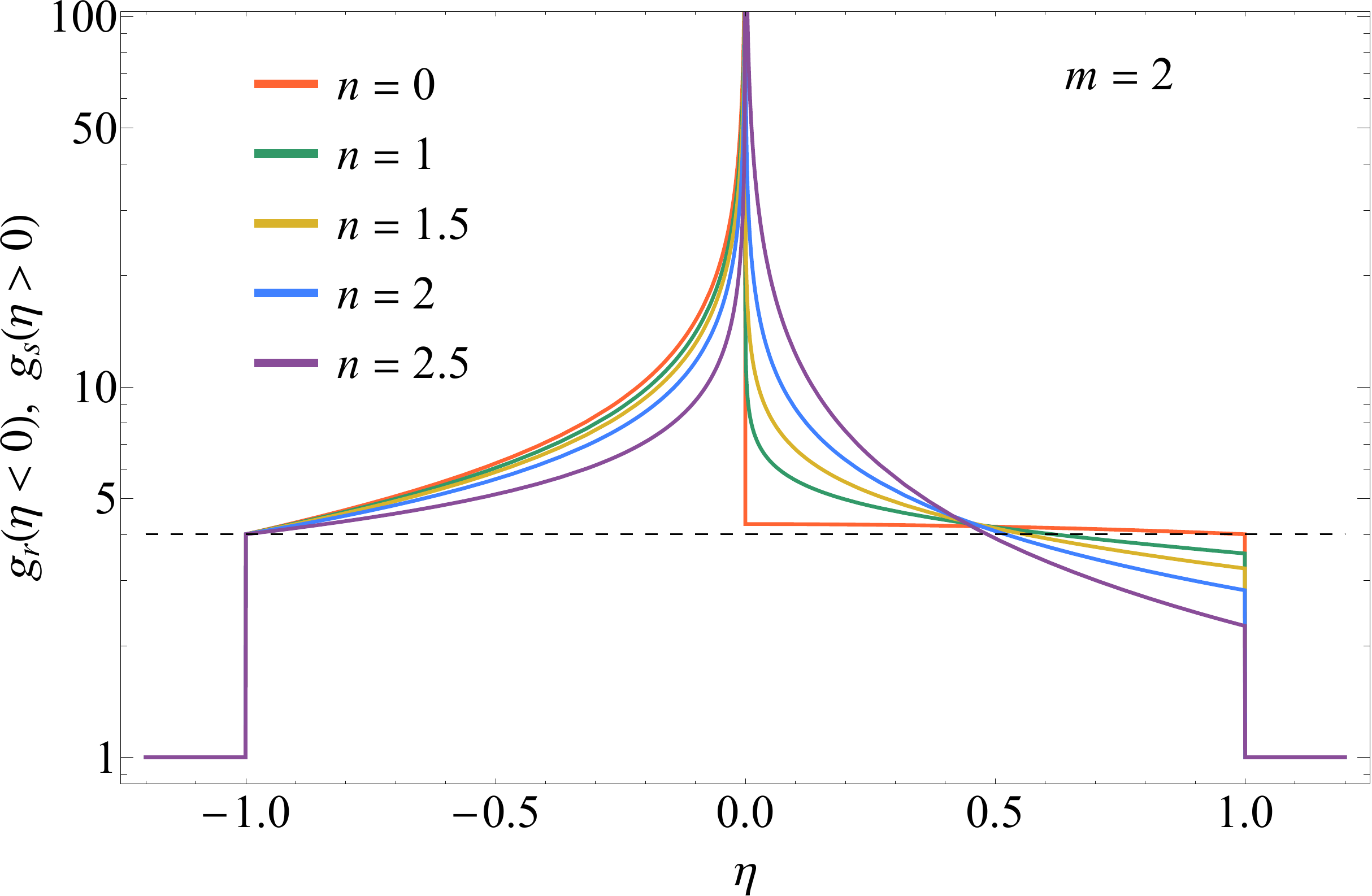} 
  \includegraphics[width=0.495\textwidth]{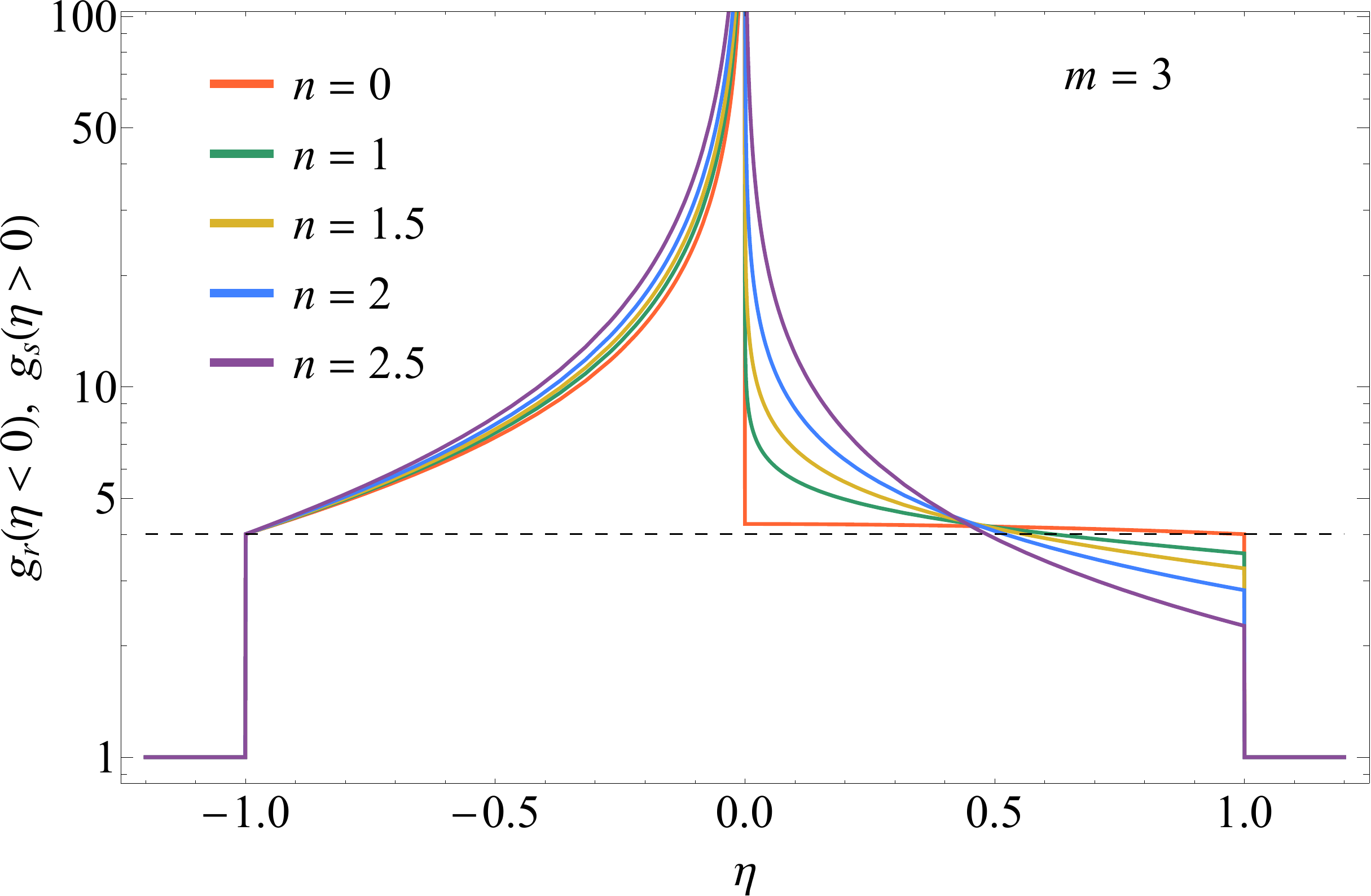} 
    \includegraphics[width=0.495\textwidth]{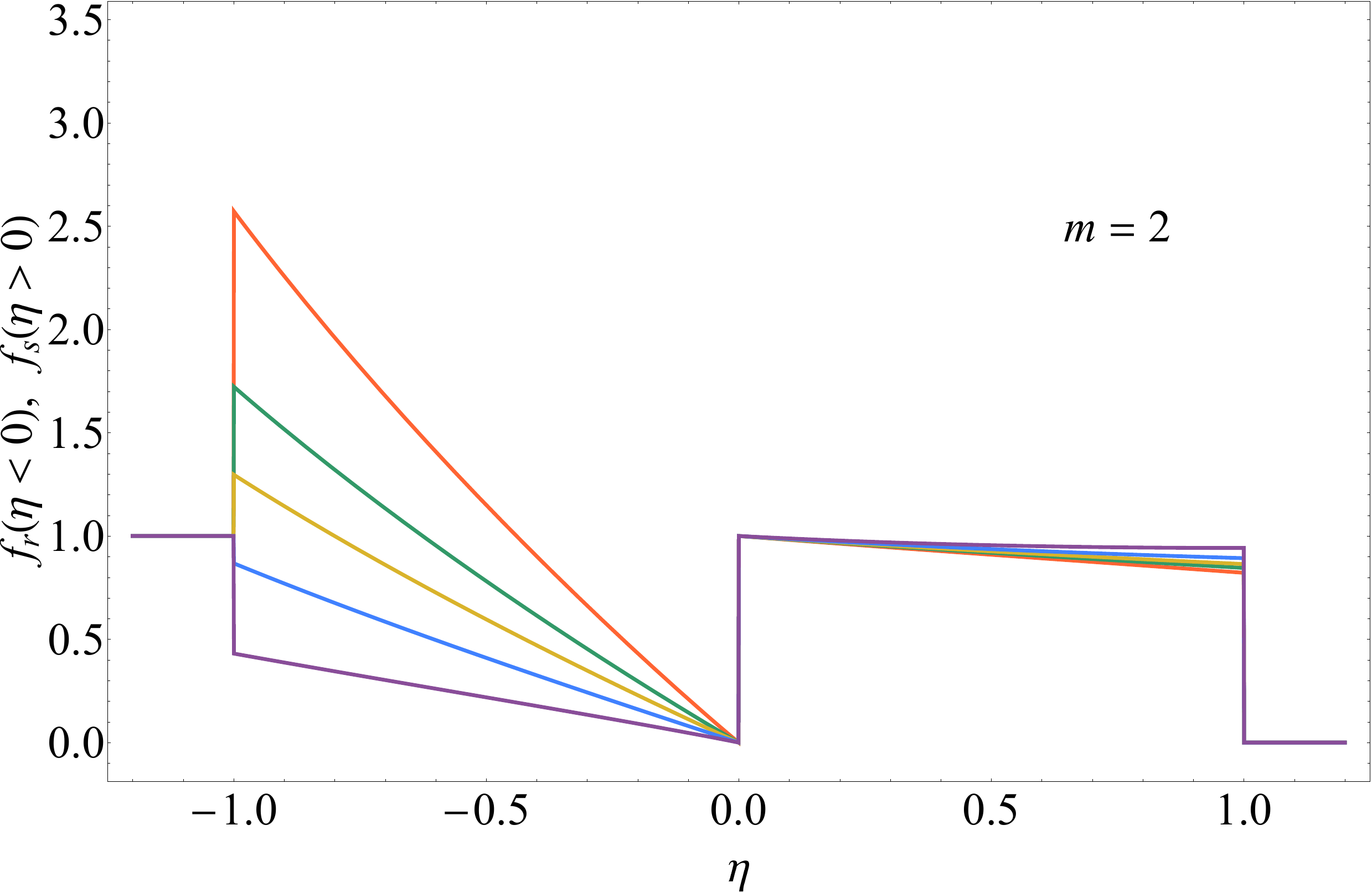} 
  \includegraphics[width=0.495\textwidth]{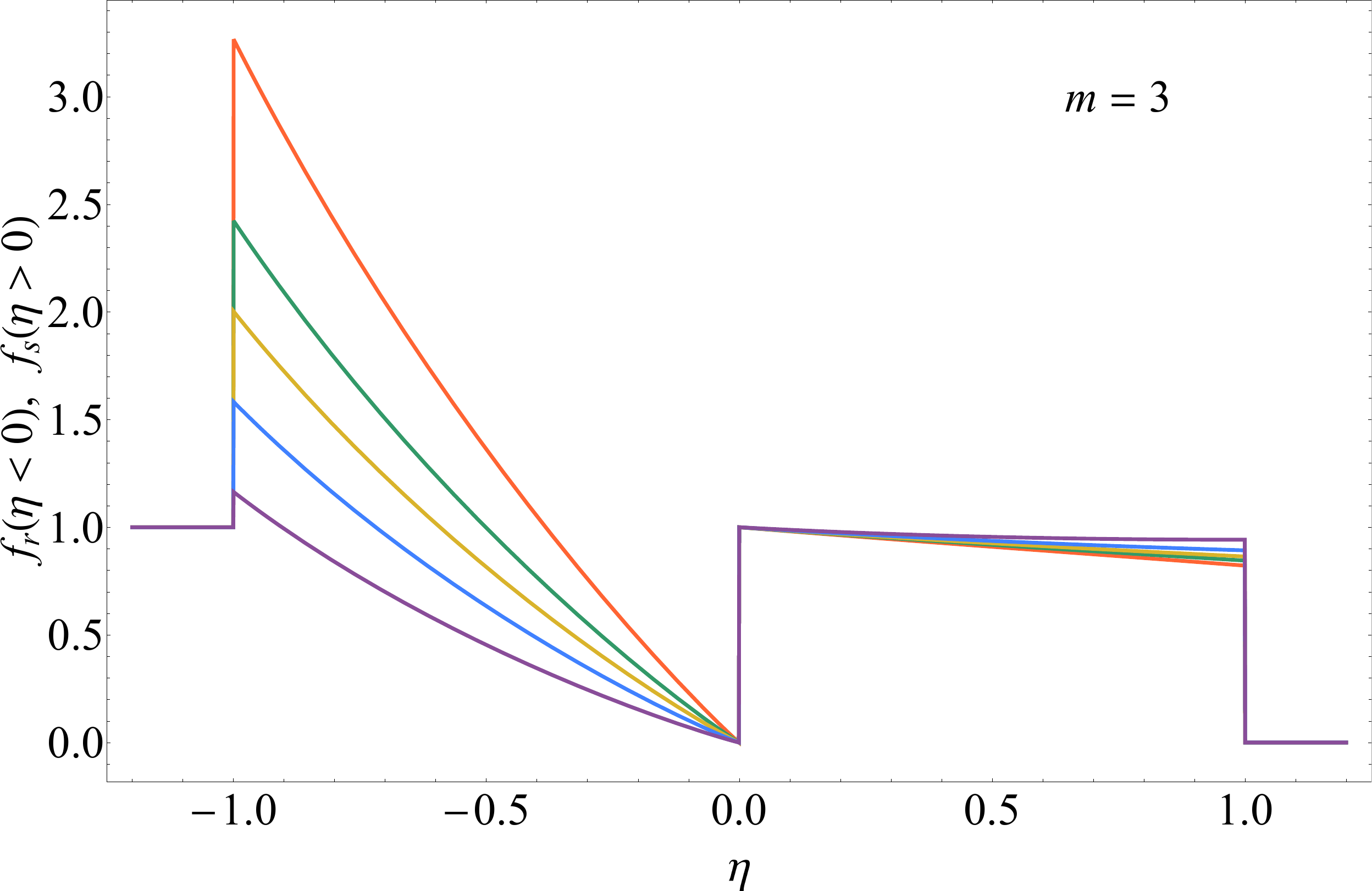} 
      \includegraphics[width=0.495\textwidth]{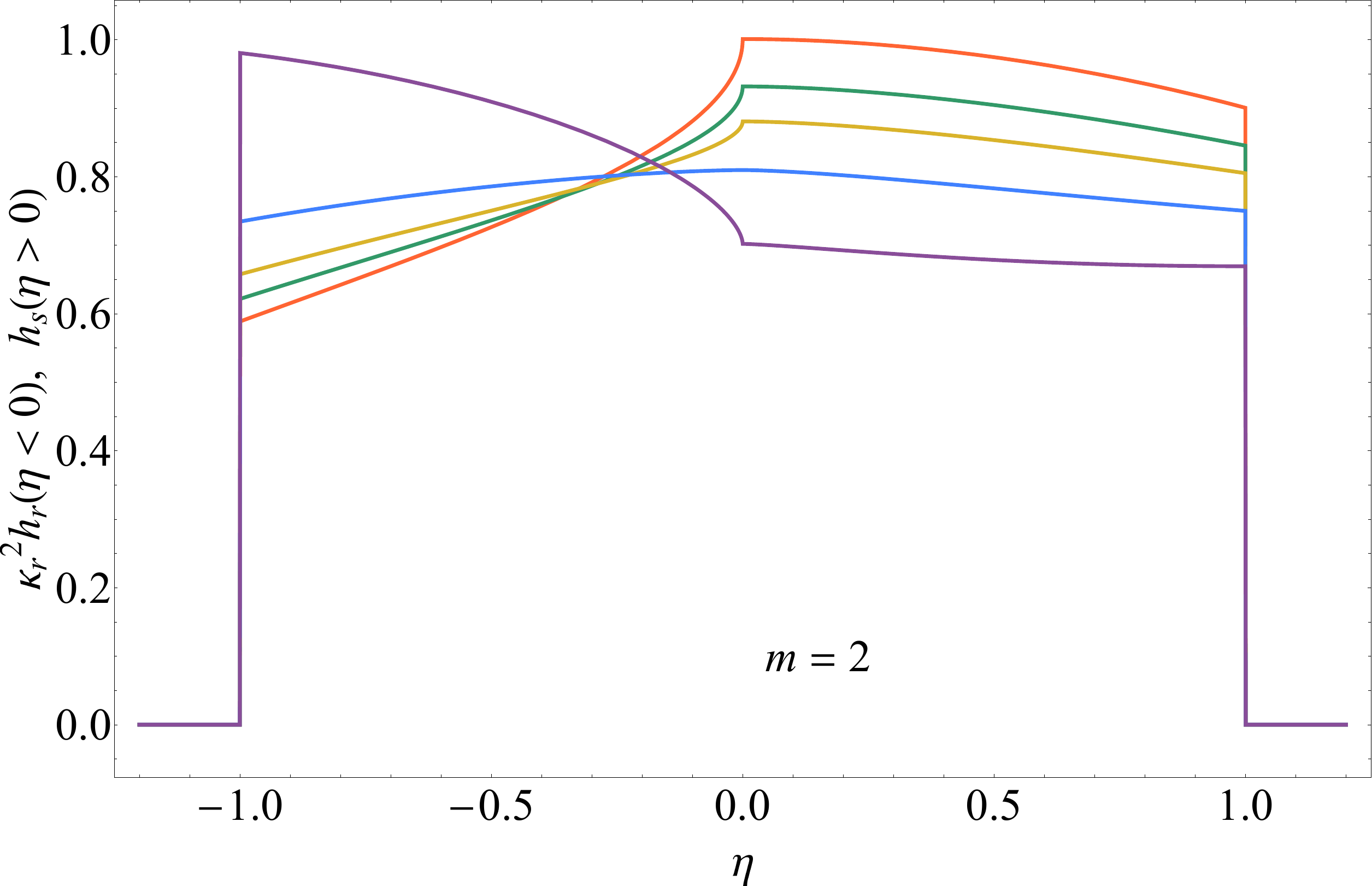} 
  \includegraphics[width=0.495\textwidth]{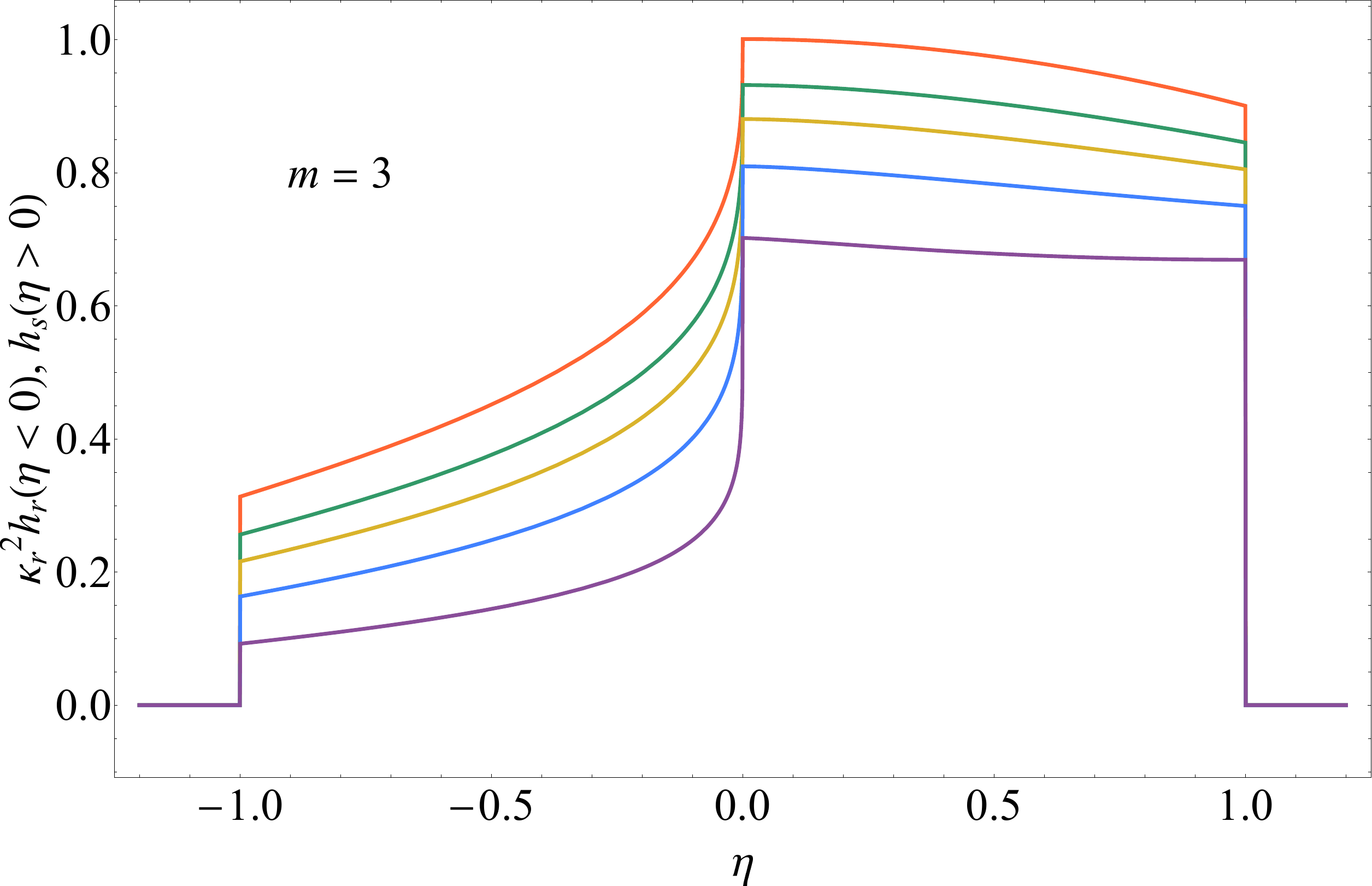} 
   \caption{The self-similar density (top), velocity (middle), and pressure (bottom) for a wind-driven explosion ($m = 2$, left) and that driven by homologous ejecta ($m = 3$, right) for the values of the ambient density power-law index $n$ shown in the legend. In all cases we set $\gamma = 5/3$. }
   \label{fig:fr_gr_hr}
\end{figure*}

With this parameterization, the boundary conditions at the FS (read off from Equation \ref{jumpbcs}) are 
\begin{equation}
\begin{split}
&f_1(1) = \frac{2}{\gamma+1}\left(1+\frac{m-n}{2}\right)\kappa_{\rm s}, \\
&g_1(1) = -\frac{\gamma+1}{\gamma-1}\left(1+\Delta_{\rm s}\right)^{-n-1}n\kappa_{\rm s}, \\
&h_1(1) = \frac{2}{\gamma+1}\left(1+\Delta_{\rm s}\right)^{1-n}\left(m+2-2n\right)\kappa_{\rm s}, \label{bcsf1}
\end{split}
\end{equation}
while the continuity of the velocity at the CD demands
\begin{equation}
f_{1}(0) = 0. \label{bcf12}
\end{equation}
Linearizing the fluid equations in $\delta$ yields the equations for $f_1$, $g_1$, and $h_1$; they are lengthy, and to preserve readability, we put them in Appendix \ref{sec:perturbations}. The solution for $\kappa_{\rm s}$ is found, as for $\kappa_{\rm c}/\kappa_{\rm r}$, by shooting from the FS until $f_{1}(0) = 0$. 

\begin{figure*}[t!] 
   \centering
   \includegraphics[width=0.495\textwidth]{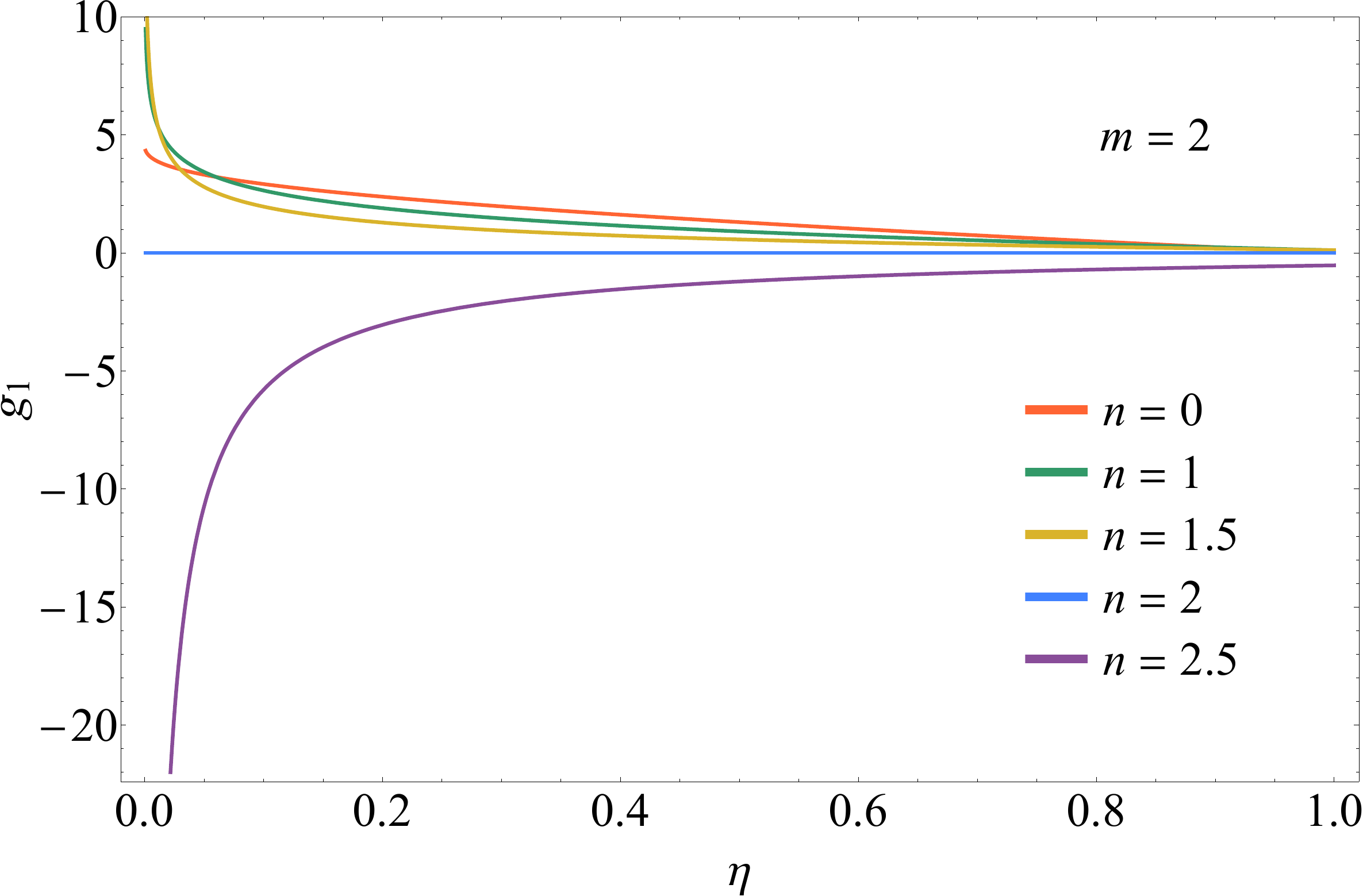} 
  \includegraphics[width=0.495\textwidth]{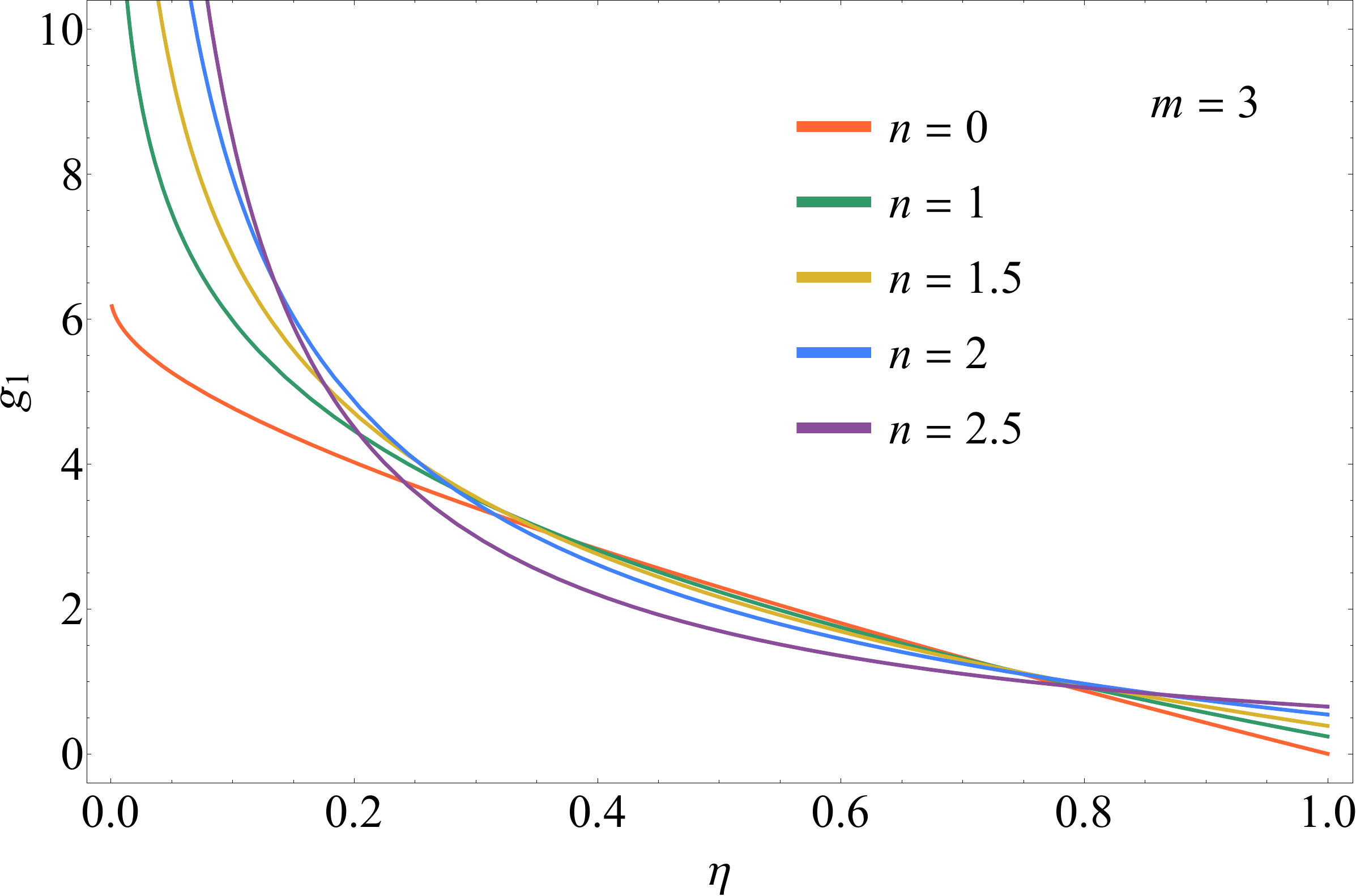} 
    \includegraphics[width=0.495\textwidth]{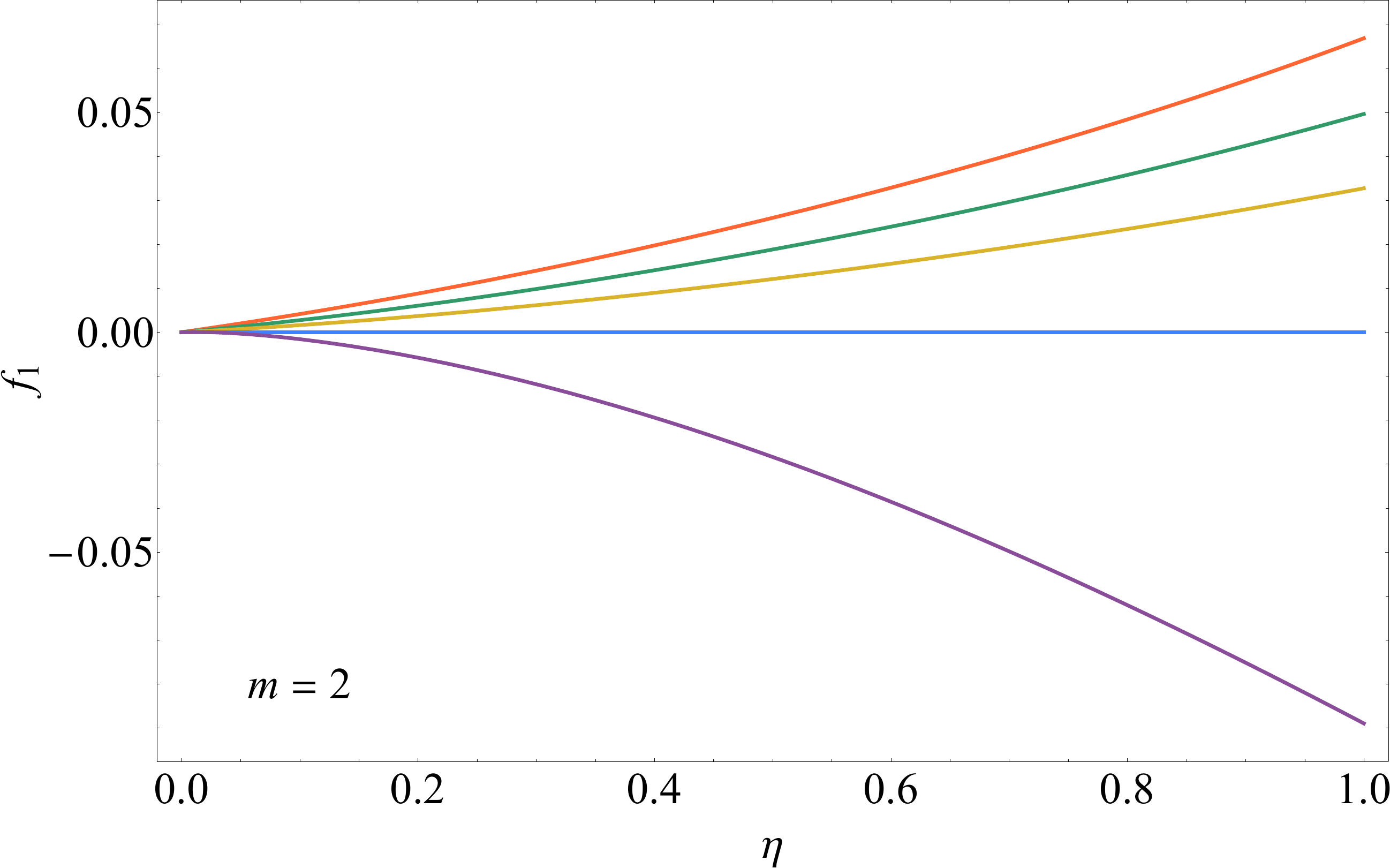} 
  \includegraphics[width=0.495\textwidth]{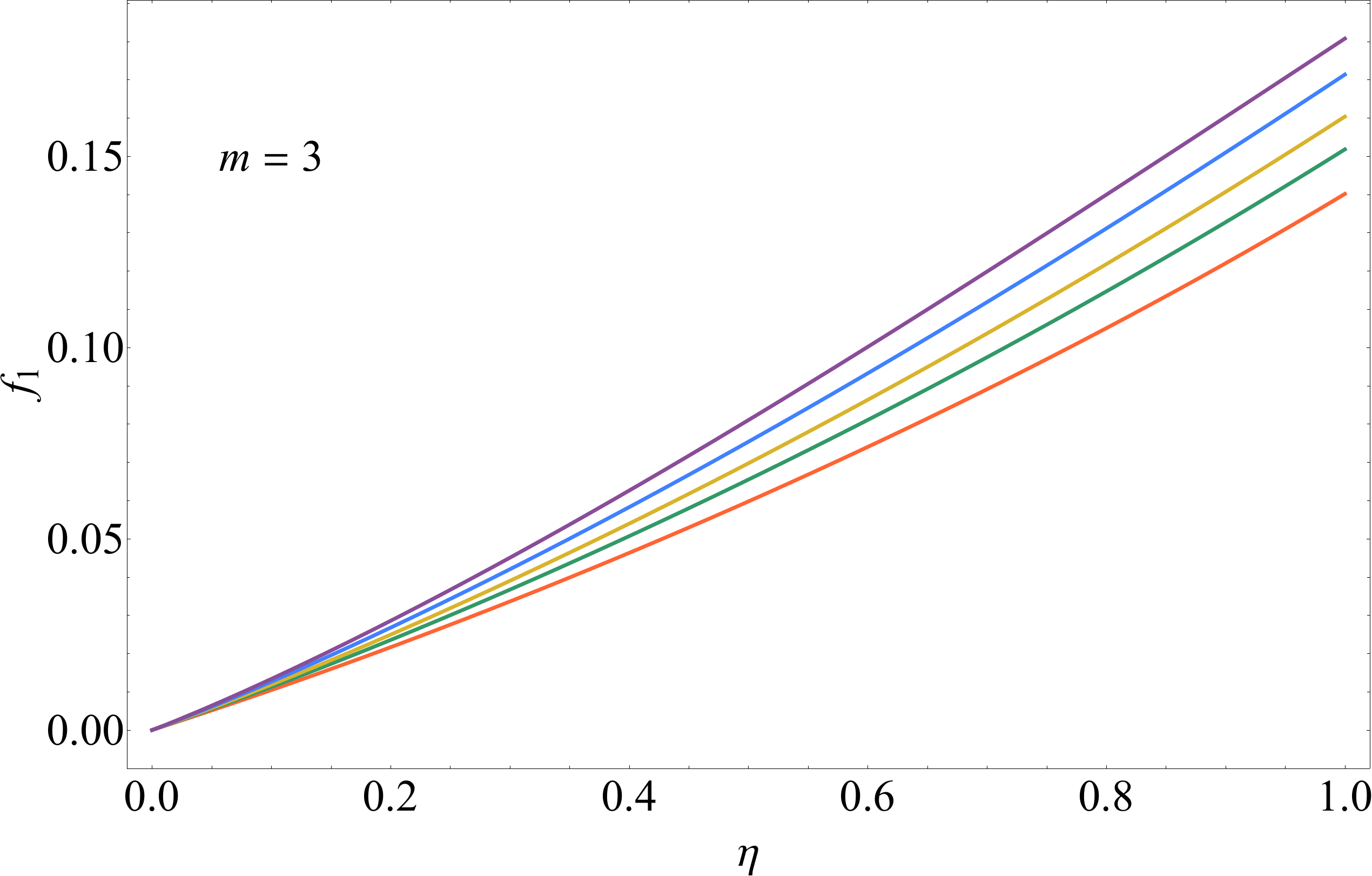}
      \includegraphics[width=0.495\textwidth]{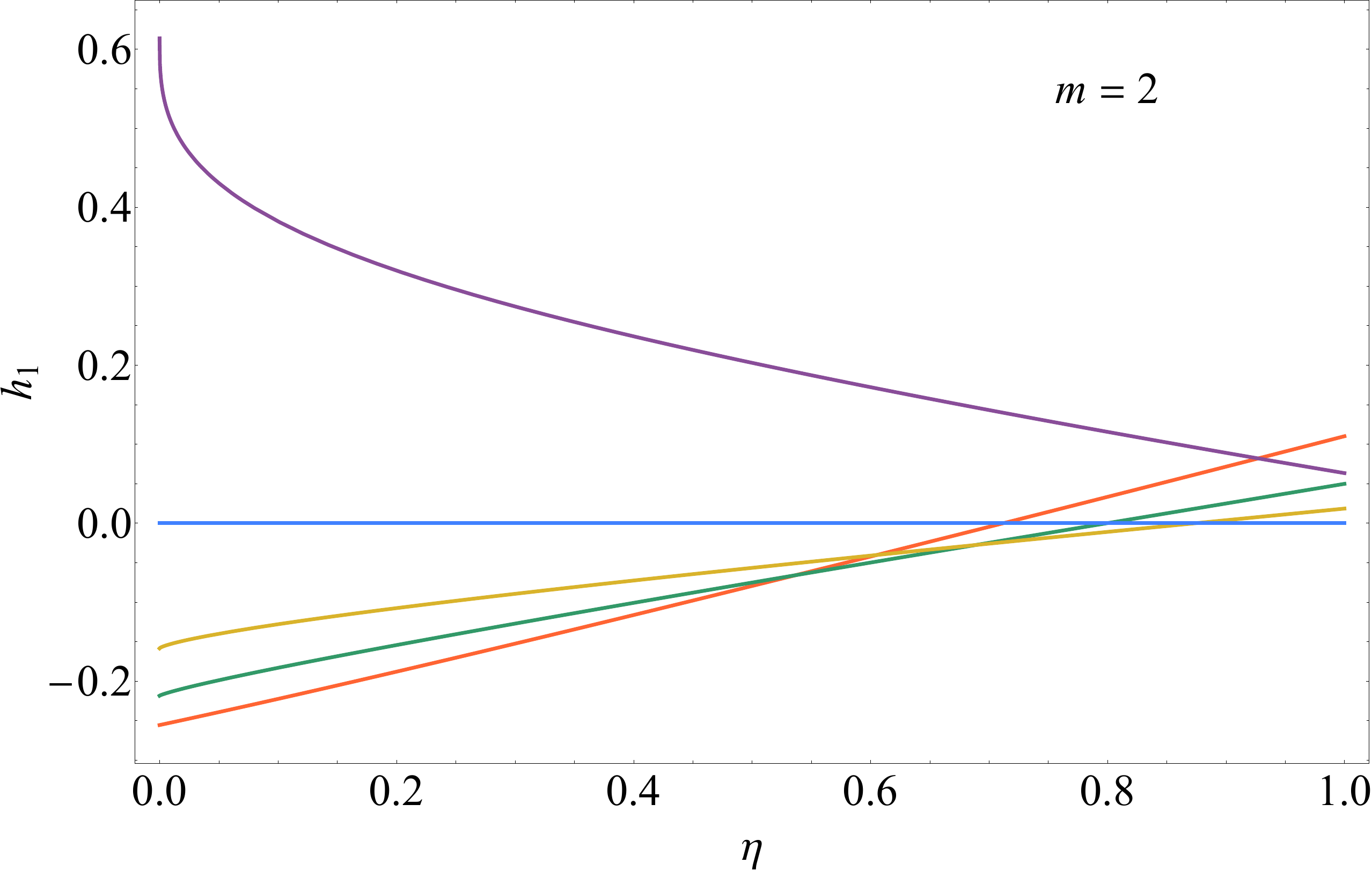} 
  \includegraphics[width=0.495\textwidth]{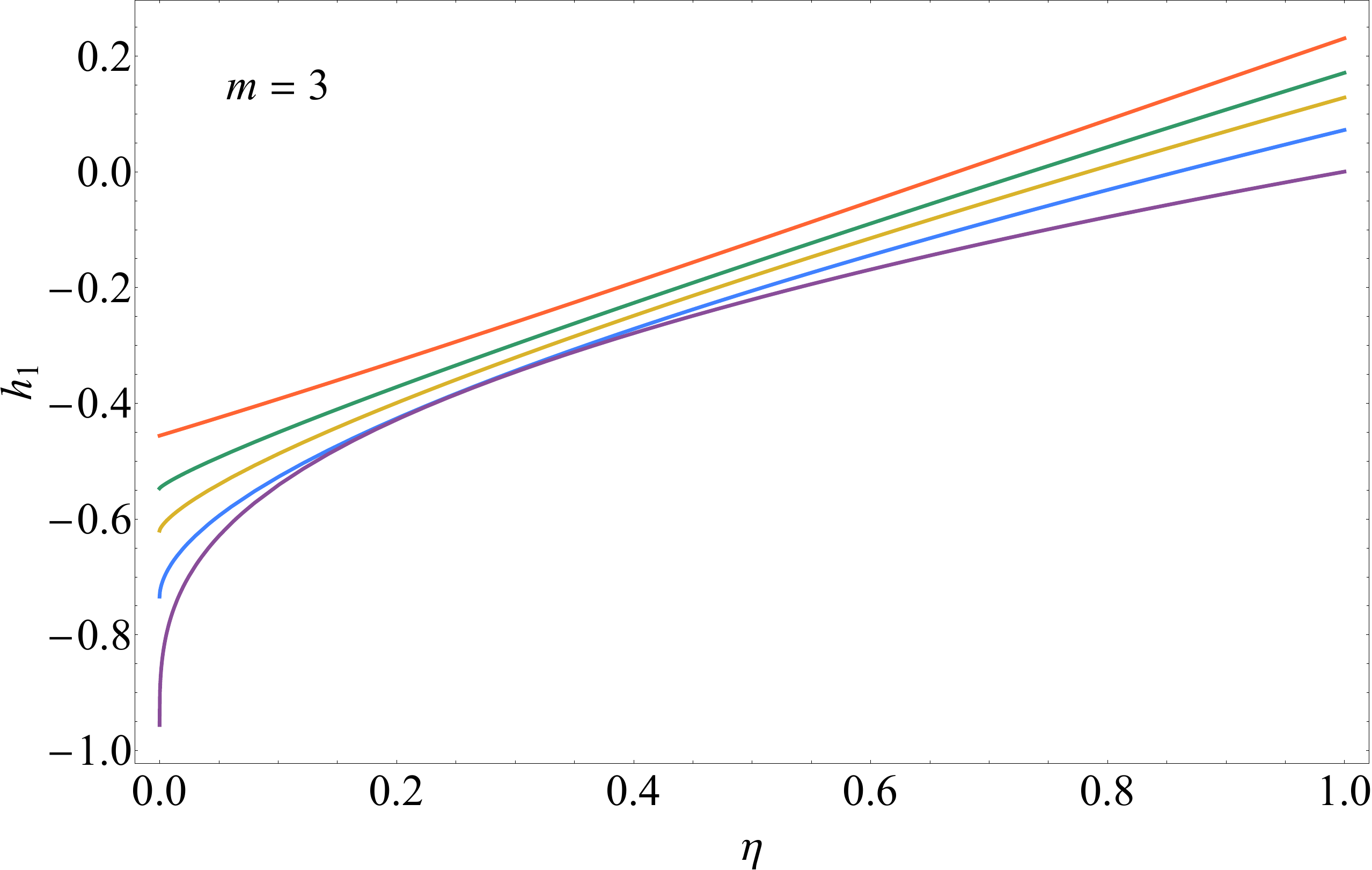} 
   \caption{Analogous to Figure \ref{fig:fr_gr_hr}, but here we are plotting the corrections to the FS density $g_1$, velocity $f_1$, and pressure $h_1$.} 
   \label{fig:f1_g1_h1}
\end{figure*}

\subsection{Total solutions}
\label{sec:total}
Figure \ref{fig:fr_gr_hr} shows the self-similar solutions for $m = 2$ (ejecta in the form of a time-steady wind; left panels) and $m = 3$ (homologously expanding ejecta; right panels) for the $n$ in the legend of the top panels and $\gamma = 5/3$; the solutions are qualitatively similar for $\gamma = 4/3$. The density (top panels) of the shocked ejecta increases and diverges at the CD for all values of $n$, such that  most of the mass is concentrated near the CD. The velocity profile is roughly linear throughout the shell, while the pressure displays nontrivial variation throughout the RS shell and is approximately uniform for the FS. Figure \ref{fig:f1_g1_h1} is analogous to Figure \ref{fig:fr_gr_hr}, but illustrates the corrections to the FS fluid variables. Table \ref{tab:1} provides values of $\Delta_{\rm s}$ and the various $\kappa$'s for both values of $m$ and a range of $n$. 

\begin{table*}
\centering
\begin{tabular}{|c|c|c|}
\hline
$\{\Delta_{\rm s}, \kappa_{\rm r}, \kappa_{\rm c}, \kappa_{\rm s}\}\times 10$ & $m =2$ & $m = 3$ \\
\hline
$n = 0$ & \hspace{-.3in} \begin{tabular}{c} \,\,\,$\gamma = 5/3$: \{0.957, 0.484, 3.95, 0.334\} \\ $\gamma = 4/3$: \{0.512, 0.266, 4.00,0.215\}\end{tabular} & \hspace{-.3in}\begin{tabular}{c} \,\,\,$\gamma = 5/3$: \{0.957, 0.280, 4.03, 0.561\} \\ \,\, $\gamma = 4/3$: \{0.595, 0.153, 4.05, 0.363\}\end{tabular}  \\
\hline
$n = 0.5$ & \hspace{-.3in} \begin{tabular}{c} \,\,\,$\gamma = 5/3$: \{1.09, 0.574, 4.54, 0.343\} \\ $\gamma = 4/3$: \{0.595, 0.325, 4.64, 0.229\}\end{tabular} & \hspace{-.3in} \begin{tabular}{c} \,\,\,$\gamma = 5/3$: \{1.09, 0.305, 4.66, 0.647\} \\ $\gamma = 4/3$: \{0.595, 0.170, 4.71, 0.433\}\end{tabular} \\
\hline
$n = 1$ & \hspace{-.3in} \begin{tabular}{c} \,\,\,$\gamma = 5/3$: \{1.27, 0.706, 5.36, 0.331\} \\ $\gamma = 4/3$: \{0.711, 0.417, 5.53, 0.232\}\end{tabular} & \hspace{-.3in} \begin{tabular}{c} \,\,\,$\gamma = 5/3$: \{1.27, 0.330, 5.51, 0.759\} \\ $\gamma = 4/3$: \{0.711, 0.191, 5.63, 0.534\}\end{tabular} \\
\hline
$n = 1.5$ & \hspace{-.3in} \begin{tabular}{c} \,\,\,$\gamma = 5/3$: \{1.52, 0.919, 6.57, 0.262\} \\ $\gamma = 4/3$: \{0.884, 0.579, 6.84, 0.198\}\end{tabular} & \hspace{-.3in} \begin{tabular}{c} \,\,\,$\gamma = 5/3$: \{1.52, 0.357, 6.80, 0.916\} \\ $\gamma = 4/3$: \{0.884, 0.216, 7.00, 0.689\}\end{tabular} \\
\hline
$n = 2$ & \hspace{-.3in} \begin{tabular}{c} \,\,\,$\gamma = 5/3$: \{1.90, 1.32, 8.57, 0\} \\ $\gamma = 4/3$: \{1.17, 0.933, 9.02, 0\}\end{tabular} & \hspace{-.3in} \begin{tabular}{c} \,\,\,$\gamma = 5/3$: \{1.90, 0.378, 8.95, 1.14\} \\ $\gamma = 4/3$: \{1.17, 0.243, 9.34, 0.955\}\end{tabular} \\
\hline
$n = 2.5$ & \hspace{-.3in} \begin{tabular}{c} \,\,\,$\gamma = 5/3$: \{2.56, 2.42, 12.8, -1.19\} \\ $\gamma = 4/3$: \{1.74, 2.19, 13.4, -1.30\}\end{tabular} & \hspace{-.3in} \begin{tabular}{c} \,\,\,$\gamma = 5/3$: \{2.56, 0.363, 13.7, 1.45\} \\ $\gamma = 4/3$: \{1.75, 0.256, 14.4, 1.46\}\end{tabular} \\
\hline
$n = 3$ & \hspace{-.3in} \begin{tabular}{c} \,\,\,$\gamma = 5/3$: \{4.11, \ldots\} \\ $\gamma = 4/3$: \{3.74, \ldots\}\end{tabular} & \begin{tabular}{c} \,\,\,$\gamma = 5/3$: \{4.11, \ldots\} \\ $\gamma = 4/3$: \{3.74, \ldots\}\end{tabular} \\
\hline
\end{tabular}
\caption{$\Delta_{\rm s}$, $\kappa_{\rm r}$, $\kappa_{\rm c}$, and $\kappa_{\rm s}$ for a time-steady wind ($m = 2$, left column) and homologous ejecta ($m = 3$, right column) for various ambient density power-law indices $n$.}
\label{tab:1}
\end{table*}

\begin{figure*}[htbp] 
   \centering
       \includegraphics[width=0.495\textwidth]{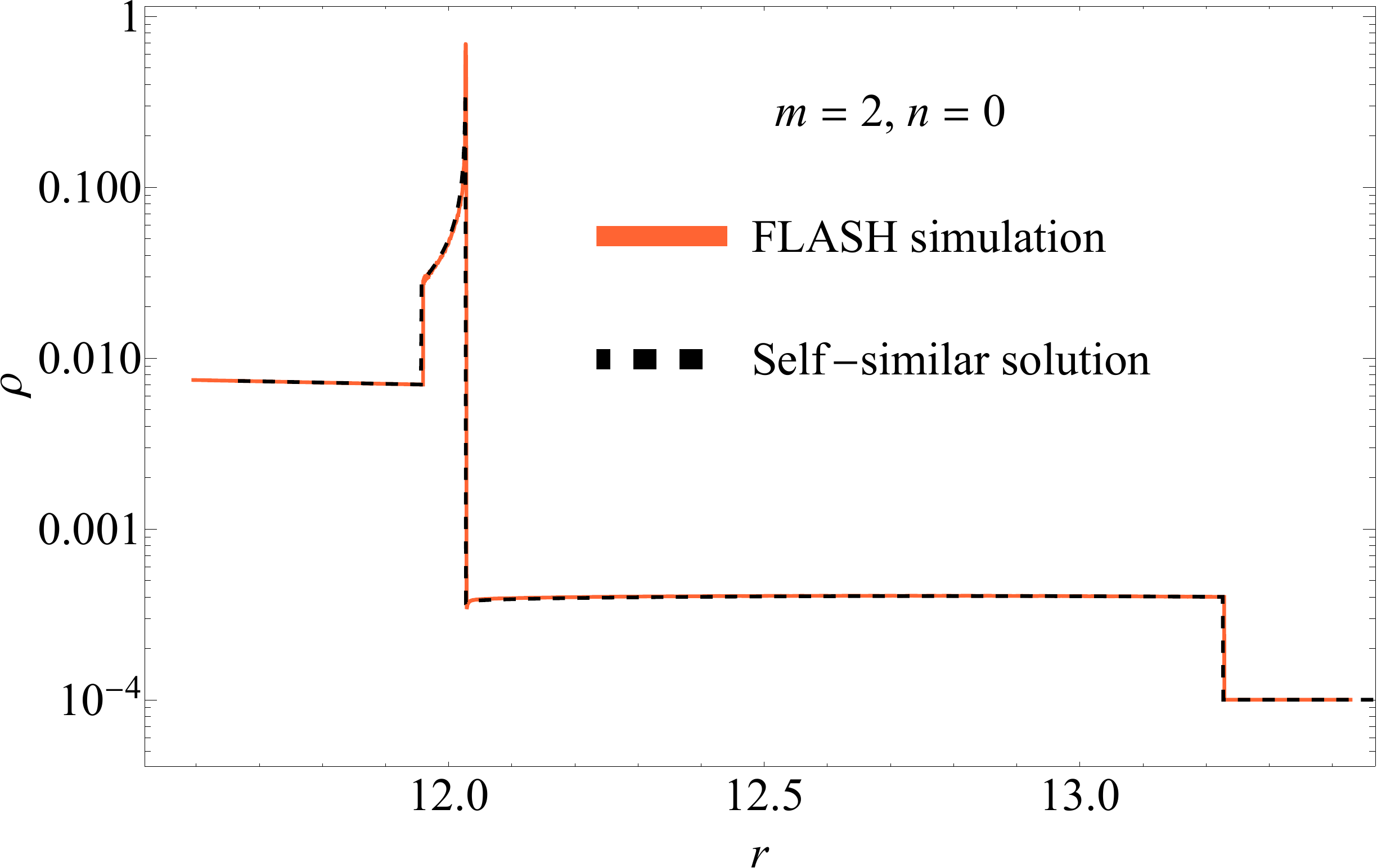} 
     \includegraphics[width=0.495\textwidth]{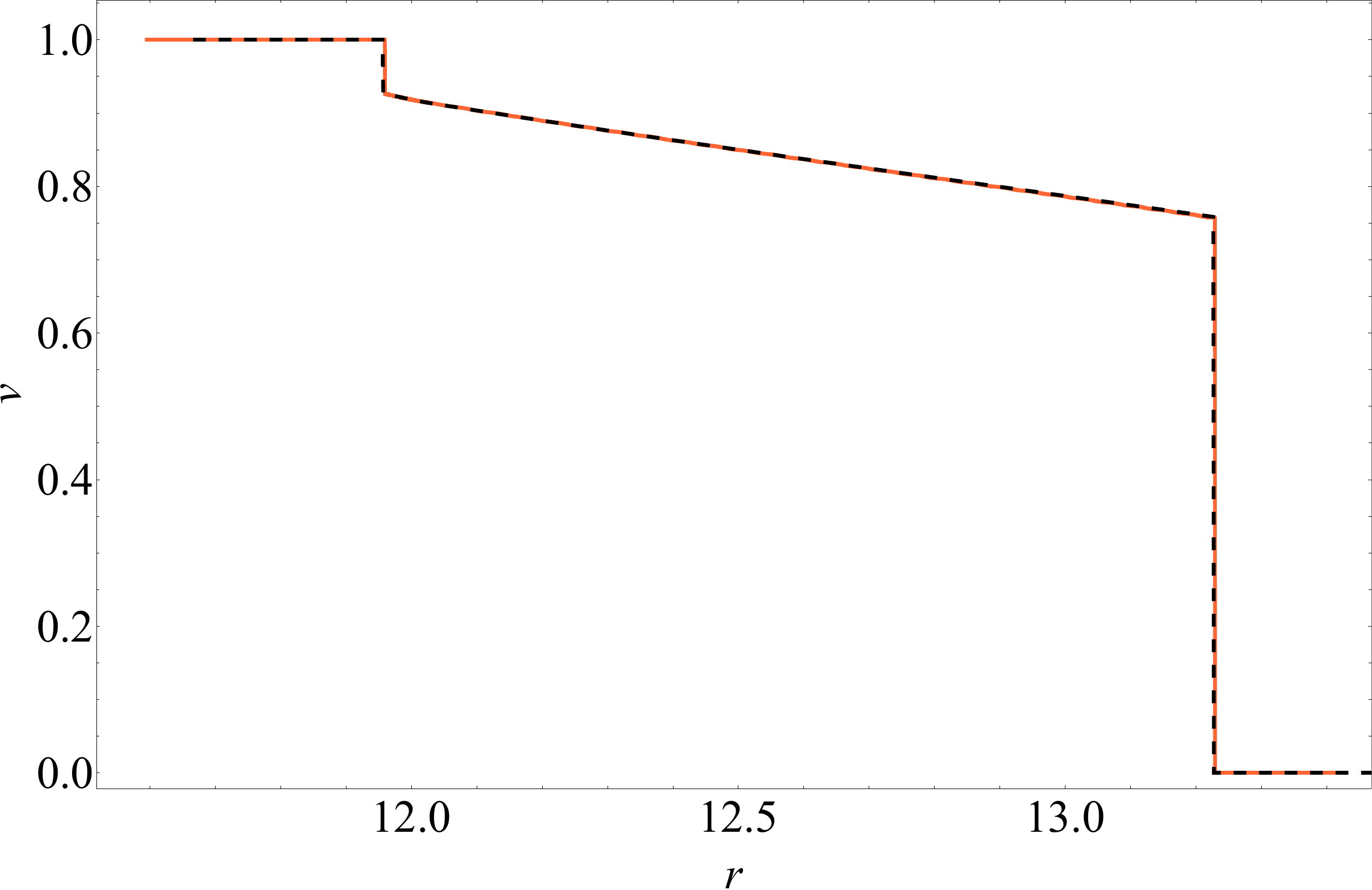} 
     \includegraphics[width=0.495\textwidth]{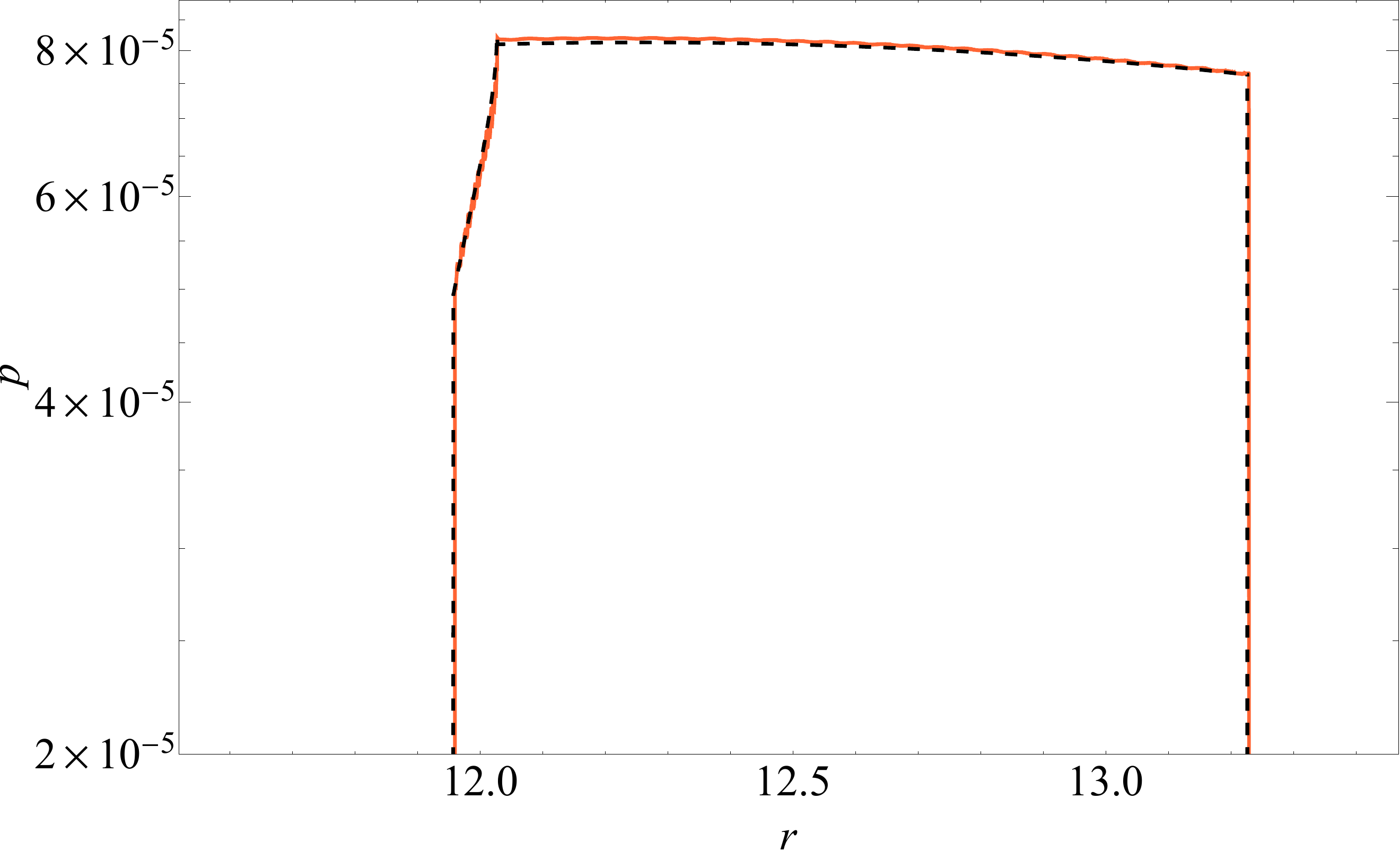} 
 \includegraphics[width=0.495\textwidth]{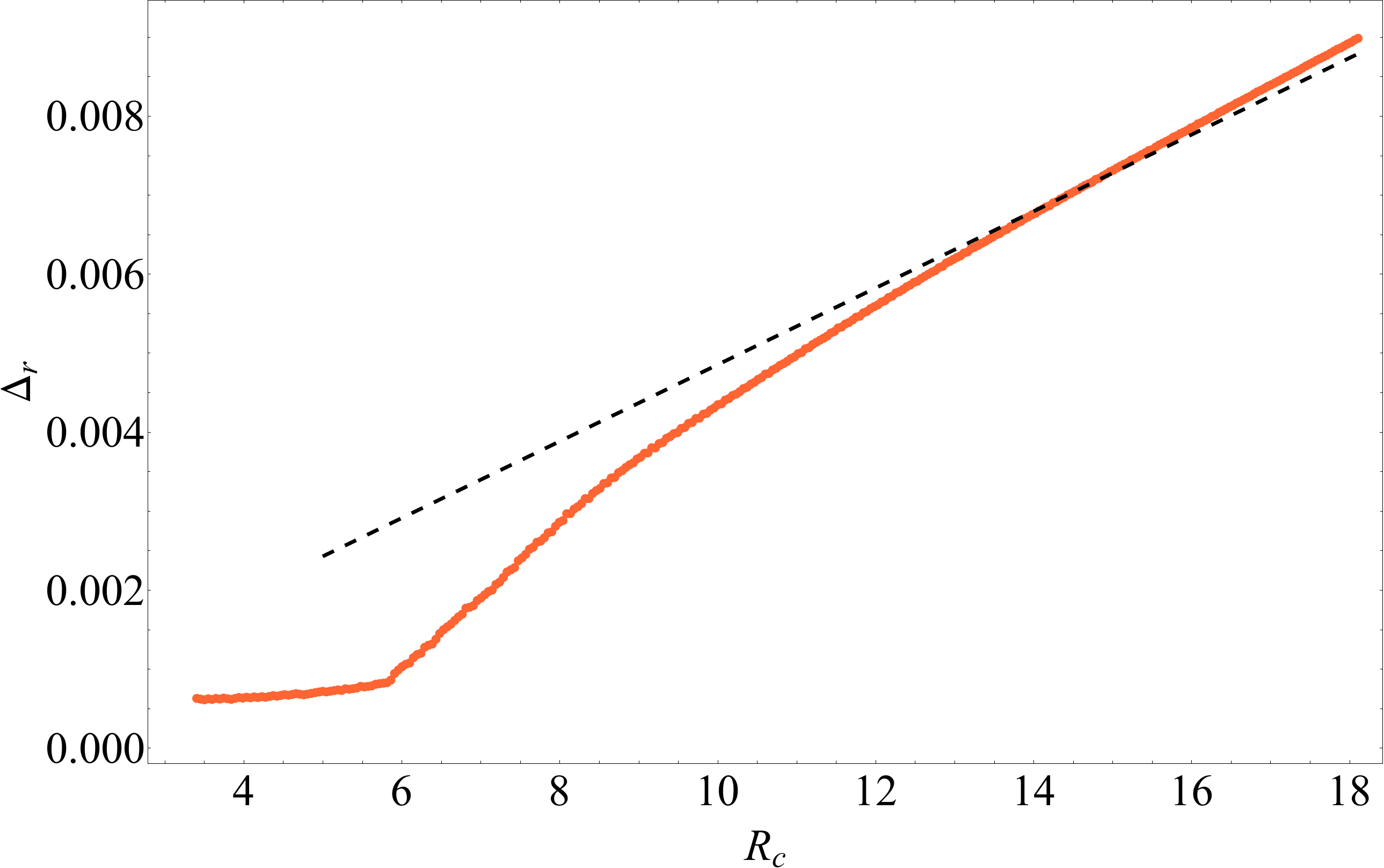} 
   \caption{The density (top-left), velocity (top-right), and pressure (bottom-left) profiles for a wind-driven explosion ($m = 2$, left) in a constant-density ($n = 0$) ambient medium and $\gamma = 5/3$. The black-dashed curves give the predictions from the self-similar solution, while the orange curves are from a {\sc flash} hydrodynamical simulation. The bottom-right panel shows the RS-shell width, $\Delta_{\rm r}$, as a function of the radius of the CD, both from {\sc flash} (orange) and as predicted by the self-similar solution (black-dashed).} 
   \label{fig:num_m2}
\end{figure*}

\begin{figure*}[htbp] 
   \centering
  \includegraphics[width=0.495\textwidth]{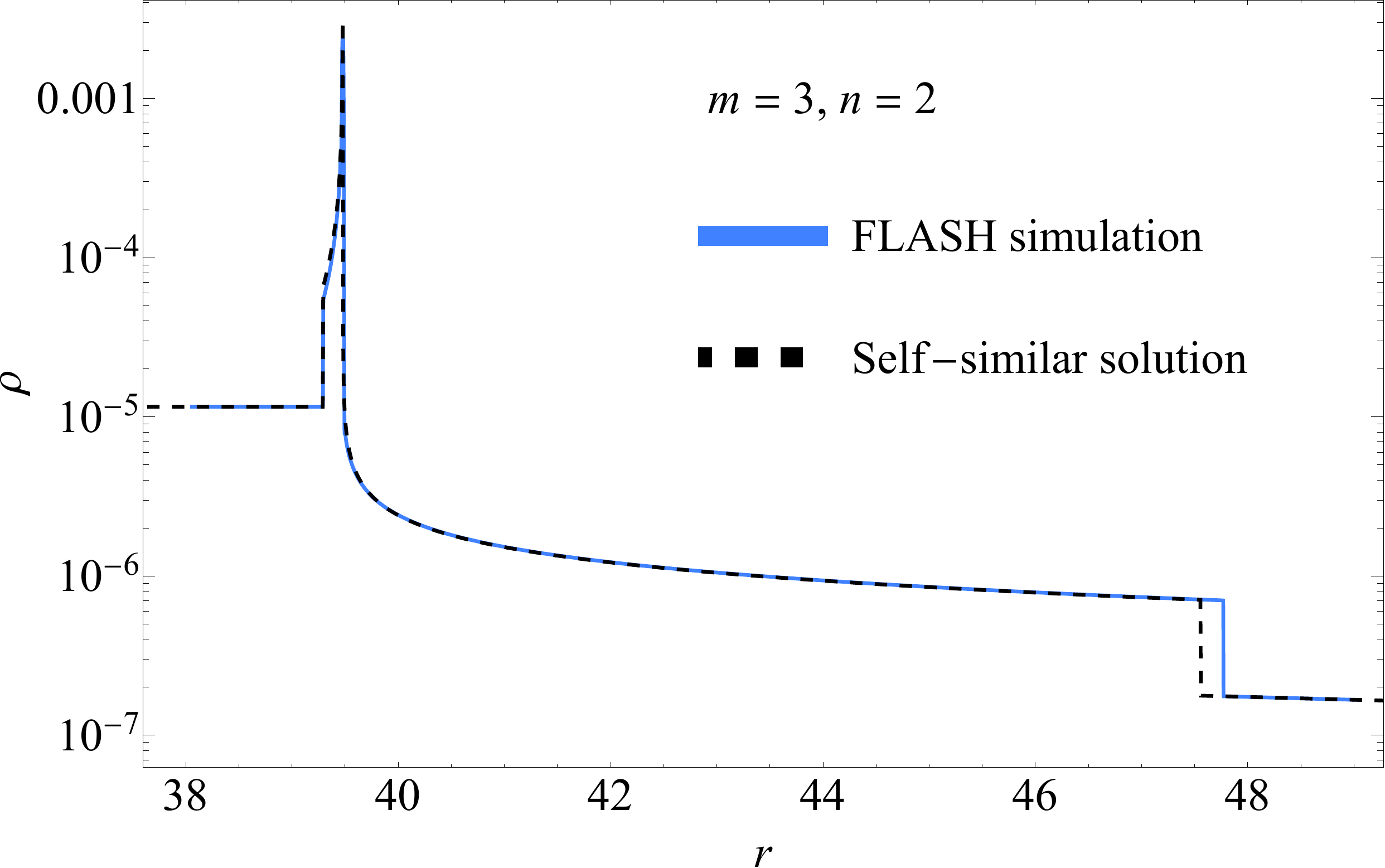} 
    \includegraphics[width=0.495\textwidth]{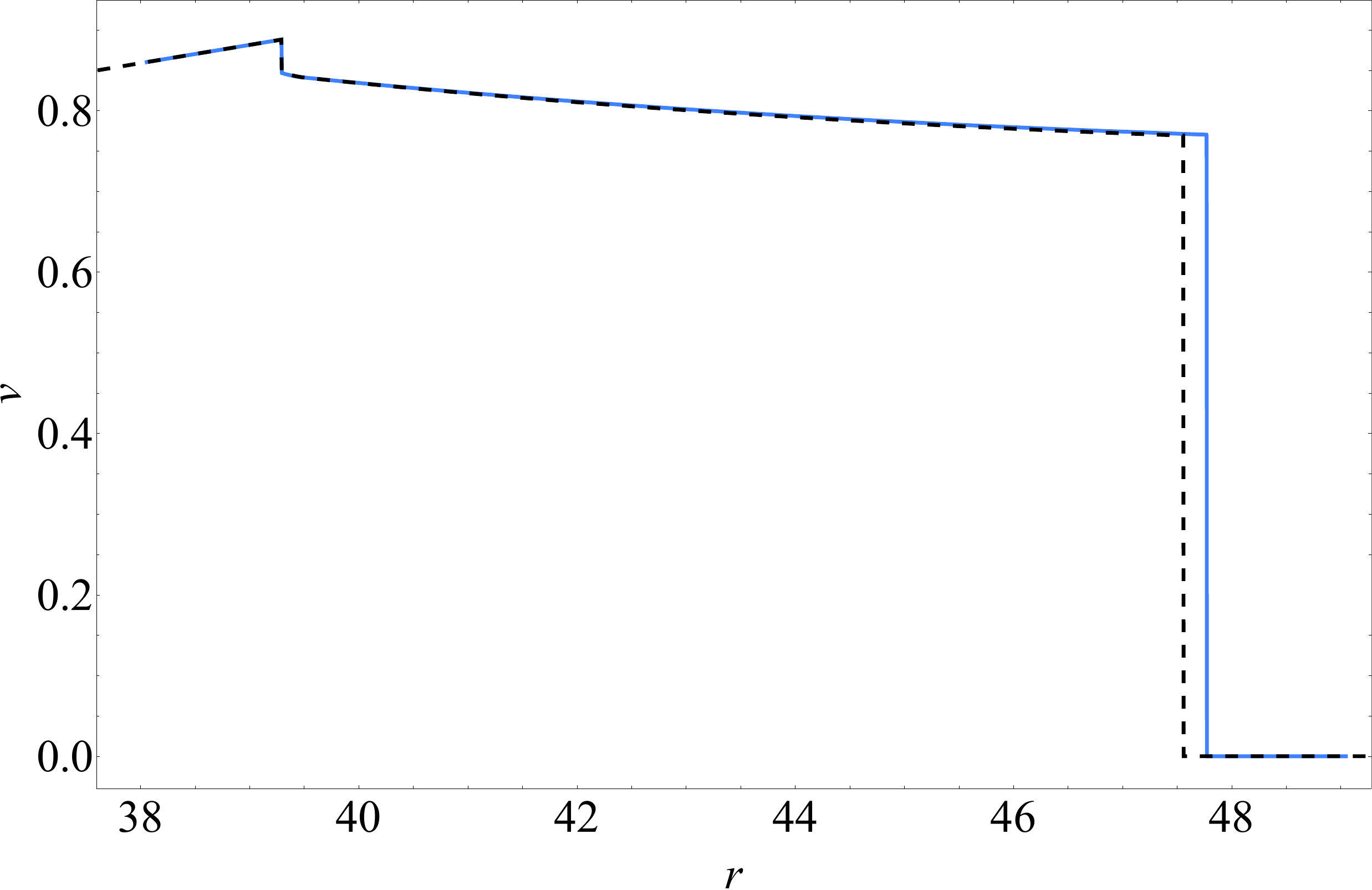} 
     \includegraphics[width=0.495\textwidth]{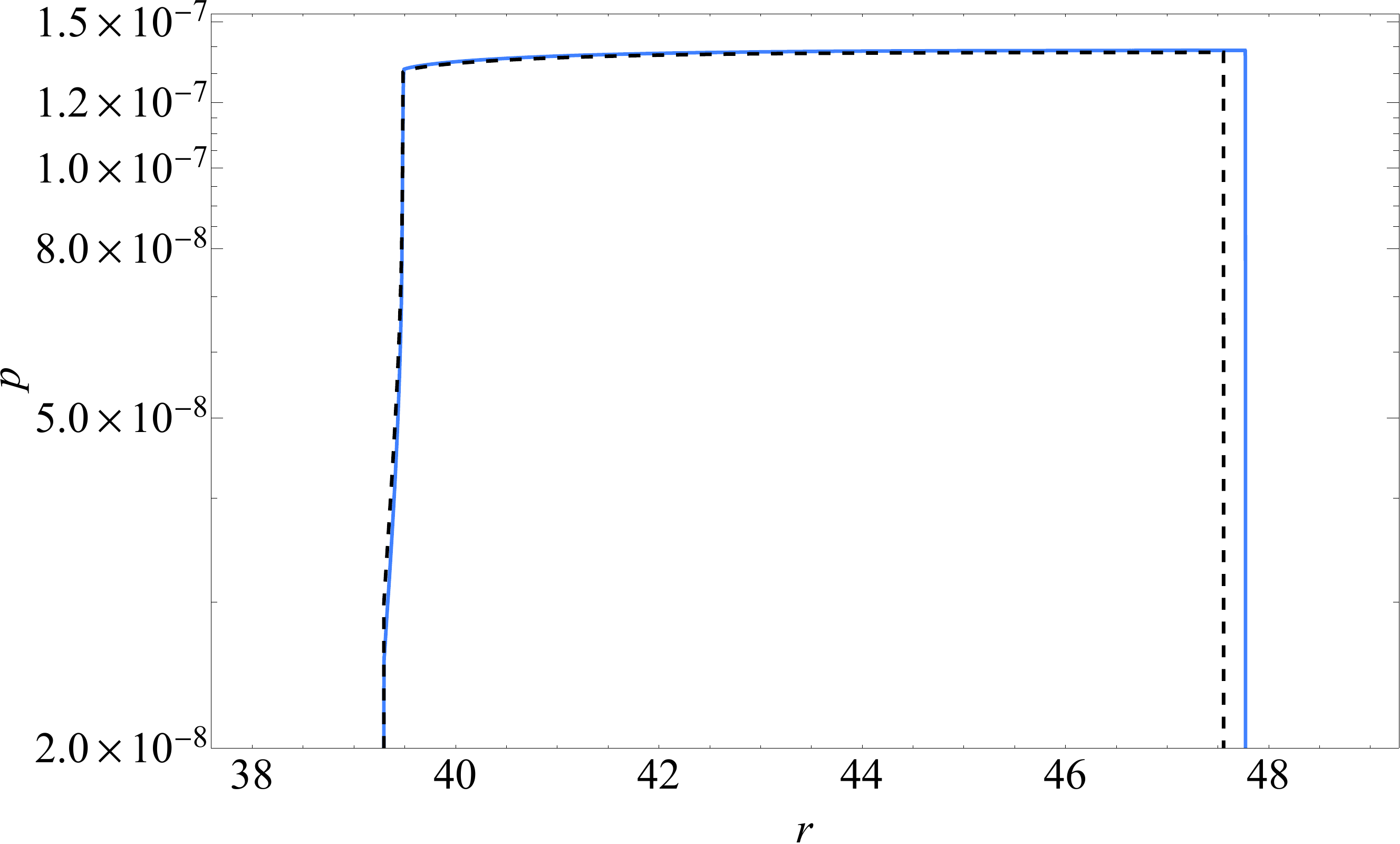} 
 \includegraphics[width=0.495\textwidth]{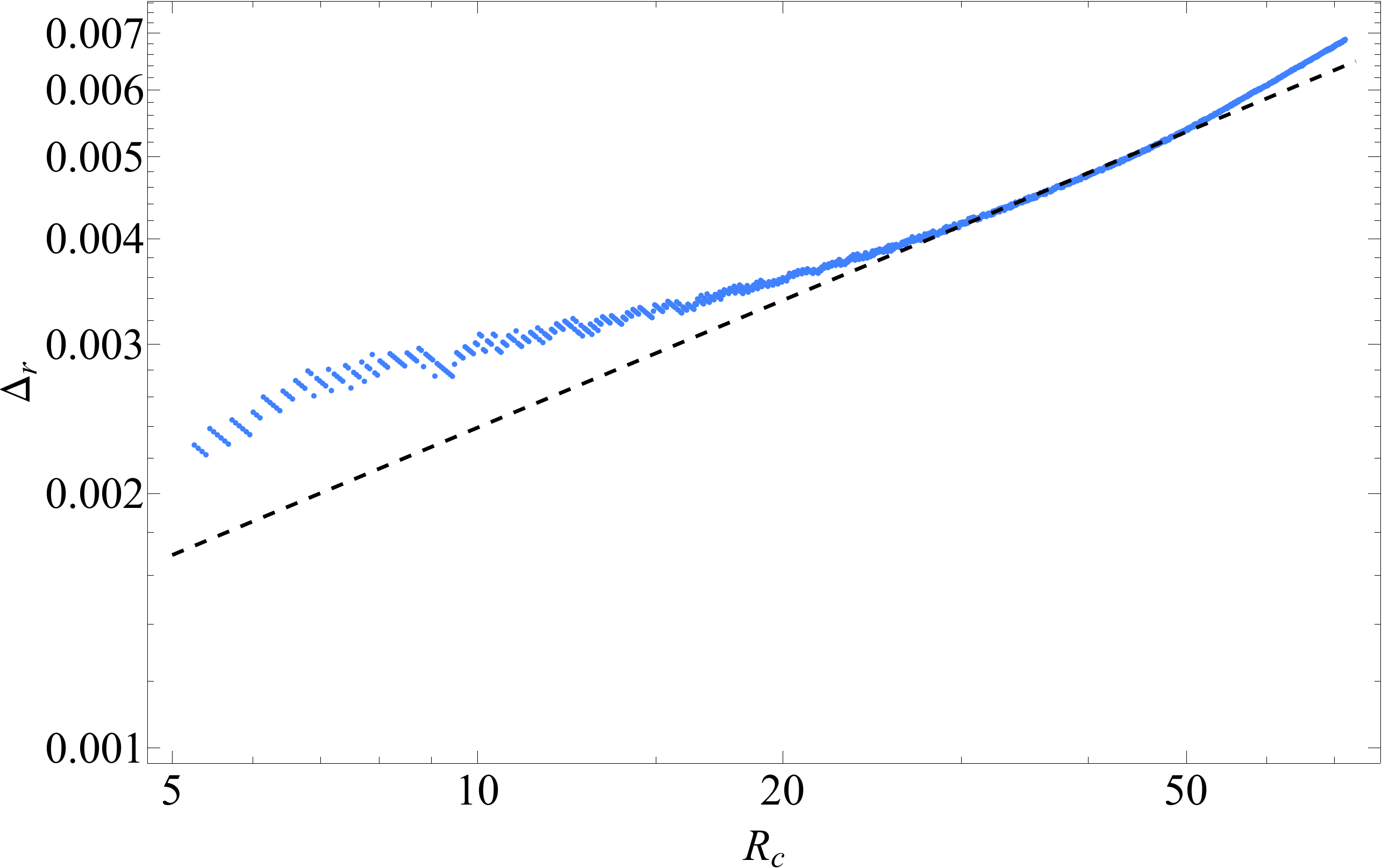} 
   \caption{Same as Figure \ref{fig:num_m2} but for homologously expanding ejecta ($m = 3$) impacting a wind-like medium ($n = 2)$. }
   \label{fig:num_m3}
\end{figure*}

The solid, orange curves in the top-left, top-right, and bottom-left panels of Figure \ref{fig:num_m2} give the density, velocity, and pressure, respectively, from a hydrodynamical simulation run with {\sc flash} \citep{fryxell00}, while the bottom-right panel shows the solution for the RS-shell width (i.e., $1-R_{\rm r}/R_{\rm c}$) as a function of the position of the CD. We initialized the simulation with a constant-velocity, cold wind, such that $v = V_{\rm ej} = 1$ and $\rho = r^{-2}$ for $r < 1$, and $v = 0$ and $\rho = 10^{-4}$ ($n = 0$) for $r > 1$, i.e., $\delta(r = 1) = 10^{-2}$. The parameters controlling numerical accuracy (e.g., interpolation order) are identical to those described in \citet{paradiso24}, and we used a resolution of $\Delta r = 7.64\times 10^{-5}$. The dashed curves are the predictions from the self-similar solutions. In all cases the agreement is extremely good. Figure \ref{fig:num_m3} is the analogous set of plots for a homologously expanding shell ($m = 3$) impacting a wind-fed medium ($n = 2$), where here the initial density ratio was set to $\rho_{\rm a}/\rho_{\rm ej} = 4\times 10^{-4}$ and the resolution $\Delta r = 10^{-3}$. For this setup the ejecta density profile was flat (independent of radius) and the initial velocity profile was $v = V_{\rm ej} r/R_{\rm ej}$, with $V_{\rm ej} = R_{\rm ej} = 1$.

\section{Discussion}
\label{sec:discussion}
Here we discuss various aspects of the self-similar solutions. 

\subsection{Duration of the self-similar phase and limiting values of $n$}
\label{sec:apply}
As for any self-similar solution, those described here do not account for initial conditions. In the idealized scenario of cold ejecta impinging on a cold medium, the initial velocities of the discontinuities are given in Equation \eqref{Vinits}, which allows us to estimate the time taken for the self-similar stage to be reached: since the initial growth rate of the FS shell is $d{\Delta}/dt = (\gamma-1)/(\gamma+1)\times V_{\rm c}/R_{\rm c}$, the self-similar solutions will be approximately valid once $\Delta \simeq \Delta_{\rm s}$, 
or in a time of
\begin{equation}
t_{\rm ss, begin} \simeq \frac{R_{\rm i}}{V_{\rm ej}}e^{\Delta_{\rm s}\frac{\gamma+1}{\gamma-1}}.
\end{equation}
For all values of $n$, $\Delta_{\rm s} \lesssim few\times 0.1$ (see Table \ref{tab:1}), meaning that the self-similar phase should be reached within only a few dynamical times of the ejecta; this is consistent with the numerical solutions in Figures \ref{fig:num_m2} and \ref{fig:num_m3}. We expect the self-similar phase to last until the RS shell $\Delta_{\rm r}$ satisfies $\Delta_{\rm r} \simeq \Delta_{\rm s}$. With $\Delta_{\rm r} = \kappa_{\rm r}\delta$, this gives
\begin{equation}
t_{\rm ss, end} \simeq \frac{R_{\rm i}}{V_{\rm ej}}\left(\frac{\Delta_{\rm s}}{\kappa_{\rm r}}\right)^{\frac{2}{m-n}}\left(\frac{\rho_{\rm ej}}{\rho_{\rm a}}\right)^{\frac{1}{m-n}}.
\end{equation}
The scaling with the ejecta and ambient density is expected on simple arguments related to the ejecta vs.~swept up mass, but since $\Delta_{\rm s}/\kappa_{\rm r} \gtrsim few$, this timescale can be considerably longer for relatively steep ambient density profiles.

We expect the FS self-similar solutions to be valid provided that the FS does not accelerate. There are second-type self-similar and accelerating solutions described in \citet{waxman93} for $n \ge n_{\rm acc}$, where $n_{\rm acc} \simeq 3.25$ for $\gamma = 5/3$ and $n_{\rm acc} \simeq 3.12$ for $\gamma = 4/3$. The RS solutions exist when $n < 3$; as $n \rightarrow 3$ the RS shell thickness goes to zero and the velocity is continuous to the CD. For $3 \le n \le n_{\rm acc}$, we expect the type-III solutions given in \citet{gruzinov03} to describe the flow. 
\subsection{Stability}
\label{sec:stability}
For $m < n$, the influence of the RS on the FS is decaying with time, meaning that the FS solution asymptotically approaches the self-similar solution. The rate at which it does so is proportional to $t^{(m-n)/2}$, provided that other perturbations affecting the solution are both stable and decay at a faster rate than $t^{(m-n)/2}$. These additional perturbations are characterized by an eigenvalue that controls the rate at which they decay (for stable solutions) or grow (for unstable solutions), and exist even in the limit that $\rho_{\rm a}/\rho_{\rm ej} \rightarrow 0$ and denote, e.g., the corrections to the shock propagation that arise from initial conditions. The equations describing the perturbations can be derived in the same way as those that arise from finite-$\delta$, except the contribution from the deceleration of the CD is ignored, and the eigenvalue $\sigma$ is determined by requiring that $f_1(0) = 0$. We have investigated these eigenvalues: all of them are stable and generally smaller than $(m-n)/2$, until the density profile of the ambient medium approaches $n_{\rm acc}$, at which point the eigenvalue approaches zero. We therefore would expect the solutions in steep density profiles ($n \sim 3$) to be most heavily modified by initial conditions, rather than the RS.  

The shocked-ejecta density diverges at the CD for all solutions (see Appendix \ref{sec:asymptotic}), seemingly implying that the RS solutions are Rayleigh-Taylor unstable. However, the deceleration of the CD and RS are themselves small in this limit, and if the Rayleigh-Taylor instability is present we expect it to grow as a power-law in $R_{\rm c}$, i.e., much more weakly than the exponential growth that accompanies a static interface. We defer a detailed investigation of the stability of the RS solutions to future work. 

\subsection{Relativistic generalizations}
\label{sec:relativistic}
When the fluid becomes relativistic, the finite speed of light {typically} implies that exact self-similar solutions cannot be found for arbitrary Lorentz factors. This is because of the inherent time dependence contained in $V_{\rm c}/c$, and only when the fluid is ultra-relativistic can self-similar solutions be found (\citealt{blandford76, best00}; but see \citealt{coughlin19}). However, this is not the case if both the shock speed and the shell thickness are constant, as the Lorentz factor  is still purely a function of the self-similar variable if we adopt the same parameterization as in Section \ref{sec:forward}. There are thus self-similar solutions for the FS and FS shell during the coasting phase for arbitrary shock speeds, spanning the non-, mildly, and ultra-relativistic regimes.

There are also analogous and relativistic solutions for the RS, but only if the condition $\Gamma_{\rm c}^2\Delta_{\rm r} \lesssim 1$ is satisfied, where $\Gamma_{\rm c} = (1-V_{\rm c}^2/c^2)^{-1/2}$. This condition effectively states that the ``natural'' shell thickness recognized by \citet{blandford76}, $R_{\rm c}/\Gamma_{\rm c}^2$, must be larger than $R_{\rm c}\Delta_{\rm r}$ for the system to be primarily determined by $\Delta_{\rm r}$. As for the Newtonian solutions, the FS ultra-relativistic generalizations are contained within the BMK space and are obtained trivially by setting $d\Gamma_{\rm c}/dt = 0$ in their self-similar equations. The RS solutions, however, are not, and are in fact dominated by the rest-mass inertia of the fluid to maintain pressure balance across the CD (see Equation \ref{ssRSdefs}). We analyze the relativistic generalizations in future work.

\subsection{Exact and higher-order solutions}
\label{sec:exact}
The self-similar solutions described here are accurate to first order in the reverse-shock shell thickness $\Delta_{\rm r}$. In general, the time-dependent nature of the shell thickness implies that exact solutions that hold for arbitrarily late times are not possible. The exception to this is if $m = n = 2$, in which case $\Delta_{\rm r}$ is a constant, and one does not need to maintain only leading-order terms in $\Delta_{\rm r}$ in the fluid equations and maintain the self-similar approximation (this is also why, in Table \ref{tab:1}, the corrections to the FS are zero for $m = n = 2$). While these solutions are interesting from an academic standpoint, the requirement of a very specific ambient power-law index makes them less practical (although a wind being driven into a wind-fed medium seems a likely astrophysical scenario), and we do not analyze them further here. 

It is also possible to extend the solutions presented here to higher-order in the shell thickness $\delta$ simply by adding more terms to the various functions, e.g., let $f_{\rm r} \rightarrow f_{\rm r}+\delta f_{1}$ in Equation \eqref{ssRSdefs}. Presumably the resulting expressions would possess a finite radius of convergence in $\delta$, which would signify the transition to the Sedov-Taylor, Weaver et al., or Chevalier self-similar solution as a function of $m$ and $n$ (and $\gamma$), analogous to the results presented in \citet{paradiso24} for the case of a shock stalling in a gravitational field.

\subsection{Time-explicit representations and examples}
\label{sec:time-explicit}
The self-similar solutions are written in terms of the CD radius $R_{\rm c}$, as this is the most obvious parameter to use, it makes the self-similar variable $\eta$ manifestly range from -1 to 1, and it yields simple boundary conditions at the RS, FS, and CD. Nonetheless, from an observational standpoint it is more useful to write the expressions for the FS and RS explicitly in terms of time, which is straightforward because $R_{\rm c}$ is, to leading self-consistent order in $\delta$, 
\begin{equation}
R_{\rm c} = R_{\rm ej}(t)\left(1-\kappa_{\rm c}\sqrt{\frac{\rho_{\rm a}}{\rho_{\rm ej}}}\left(\frac{R_{\rm ej}(t)}{R_{\rm i}}\right)^{\frac{m-n}{2}}\right),
\end{equation}
where, again, 
\begin{equation}
R_{\rm ej}(t) = R_{\rm i}\left(1+\frac{V_{\rm ej}t}{R_{\rm i}}\right).
\end{equation}
Then the FS and RS radii are
\begin{equation}
\begin{split}
&R_{\rm r} = R_{\rm ej}(t)\left(1-\left(\kappa_{\rm c}+\kappa_{\rm r}\right)\sqrt{\frac{\rho_{\rm a}}{\rho_{\rm ej}}}\left(\frac{R_{\rm ej}(t)}{R_{\rm i}}\right)^{\frac{m-n}{2}}\right), \\
&R_{\rm s} = R_{\rm ej}(t)\left(1+\Delta_{\rm s}\right) \\ 
&\times\left(1+\left(\frac{\kappa_{\rm s}}{1+\Delta_{\rm s}}-\kappa_{\rm c}\right)\sqrt{\frac{\rho_{\rm a}}{\rho_{\rm ej}}}\left(\frac{R_{\rm ej}(t)}{R_{\rm i}}\right)^{\frac{m-n}{2}}\right). \label{timeexplicit}
\end{split}
\end{equation}

As an example of the application and implications of these solutions, many TDEs have recently been found to exhibit late-time radio emission \citep{horesh21, cendes24}, which can be plausibly attributed to winds and outflows launched from the TDE and interacting with the circumnuclear medium (e.g., \citealt{matsumoto21, hayasaki23, matsumoto24}). If the mass outflow rate is even $1\%$ of the mass supply rate to the SMBH and the wind is launched at $\sim 0.1 c$ from tens of gravitational radii, the outflow density should be far in excess of the CNM density, i.e., $\rho_{\rm ej} \gg \rho_{\rm a}$. With an ambient density profile as steep as $\propto r^{-2.5}$ \citep{alexander20}, the RS shell thickness decreases with time (assuming the velocity profile of the ejecta is $\sim$ const.) as $\Delta_{\rm r}\propto t^{-1/4}$, while the FS shell thickness would be $\simeq 0.25$ (assuming $\gamma = 5/3$, or $\sim 0.17$ for $\gamma = 4/3$; see Table \ref{tab:1}). As described by \citet{matsumoto24}, the density of the CNM could flatten considerably (with $n \simeq 0$) near the sphere of influence of the SMBH, implying that at this point the RS shell thickness would grow approximately linearly with time. 

As another example, the fast blue optical transient (FBOT; \citealt{drout14}) CSS-161010 \citep{coppejans20} was interpreted as a ``dirty fireball,'' i.e., a high-mass and baryon-rich explosion powering a blastwave through a dense circumstellar medium. Because of the large ejecta density, it is possible that the system was -- even hundreds of days post-explosion -- still in the phase described here. If the ambient medium was wind-fed ($n = 2$) and the ejecta was homologously expanding ($m = 3$), we would expect the CD (where most of the mass is contained) to evolve with time as
\begin{equation}
R_{\rm c} = R_{\rm ej}(t)\left(1-0.895\sqrt{\frac{\rho_{\rm a}}{\rho_{\rm ej}}}\left(\frac{R_{\rm ej}(t)}{R_{\rm i}}\right)^{1/2}\right).
\end{equation} 
These solutions should also be applicable to other FBOTs, including AT2018cow \citep{margutti19, perley19} and ZTF18abvkwla \citep{ho20}. 

\section{Summary}
\label{sec:summary}
We analyzed new self-similar solutions to the interaction phase of both wind-driven and homologously expanding ejecta-driven explosions, for which the forward shock, reverse shock, and contact discontinuity propagate at distinct temporal rates (see Equation \ref{timeexplicit} and Table \ref{tab:1}). While these solutions are most applicable when the density of the ejecta is larger than that of the ambient medium, we expect them to provide reasonable estimates of the blastwave evolution until the Sedov-Taylor stage is reached (or the \citealt{weaver77} solution if energy is continuously injected in the form of a wind) or, if the ejecta has a steep power-law density profile, the \citealt{chevalier82} stage where all three discontinuities propagate with the same power-law dependence. These solutions therefore interpolate between the coasting and energy-conserving phases of strong explosions and apply to a wide range of astrophysical phenomena (see those described in Section \ref{sec:intro} and \ref{sec:time-explicit}). 

\section*{}
I thank the anonymous referee for useful comments and suggestions. I acknowledge support from NASA through the Astrophysics Theory Program, grant 80NSSC24K0897. 
\clearpage
\appendix

\section{Equations for the corrections to the forward shock}
\label{sec:perturbations}
The equations describing the corrections to the FS that arise from finite values of $\delta$ can be derived by letting $f_{\rm s}(\eta) \rightarrow f_{\rm s}(\eta)+\delta f_1(\eta)$ etc.~in Equation \eqref{fluiddefs}, inserting the result into the fluid equations, and keeping first-order terms in $\delta$. The result is 
\begin{multline}
\Delta_{\rm s}\frac{m-n}{2}\frac{g_1}{g_{\rm s}}-n\kappa_{\rm s}+\left(f_{\rm s}-1-\Delta_{\rm s}\eta\right)\frac{d}{d\eta}\left[\frac{g_1}{g_{\rm s}}\right]+\left(f_1-\left(1+\frac{m-n}{2}\right)\kappa_{\rm s}\eta\right)\frac{d}{d\eta}\ln g_{\rm s} \\
+\frac{df_1}{d\eta}+\frac{2\Delta_{\rm s}}{1+\Delta_{\rm s}\eta}f_1+\frac{2\kappa_{\rm s}}{\left(1+\Delta_{\rm s}\eta\right)^2}f_{\rm s} = 0,
\end{multline}
\begin{equation}
\Delta_{\rm s}\frac{m-n}{2}\left(f_1-\kappa_{\rm c}\left(1+\frac{m-n}{2}\right)\right)+\left(f_{\rm s}-1-\Delta_{\rm s}\eta\right)\frac{df_1}{d\eta}+\left(f_1-\left(1+\frac{m-n}{2}\right)\kappa_{\rm s}\eta\right)\frac{df_{\rm s}}{d\eta}+\frac{1}{g_{\rm s}}\frac{dh_1}{d\eta}-\frac{g_1}{g_{\rm s}^2}\frac{dh_{\rm s}}{d\eta} = 0,
\end{equation}
\begin{multline}
\Delta_{\rm s}\frac{m-n}{2}\left(\frac{h_1}{h_{\rm s}}-\frac{\gamma g_1}{g_{\rm s}}-2\kappa_{\rm c}\left(1+\frac{m-n}{2}\right)\right)+n\left(\gamma-1\right)\kappa_{\rm s}+\left(f_{\rm s}-1-\Delta_{\rm s}\eta\right)\frac{d}{d\eta}\left[\frac{h_1}{h_{\rm s}}-\frac{\gamma g_1}{g_{\rm s}}\right] \\ 
+\left(f_1-\left(1+\frac{m-n}{2}\right)\kappa_{\rm s}\eta\right)\frac{d}{d\eta}\ln\left(\frac{h_{\rm s}}{g_{\rm s}^{\gamma}}\right) = 0.
\end{multline}
These are solved alongside the boundary conditions \eqref{bcsf1} -- \eqref{bcf12}. 

\section{Behavior of the reverse-shock solutions near the contact discontinuity}
\label{sec:asymptotic}
The boundary condition needed to determine the deceleration rate of the RS and the CD, Equation \eqref{bcp}, necessitates that the pressure remain finite at the CD. To demonstrate that this is the case, we write the leading-order (in $\eta$) expansions of the functions $f_{\rm r}$, $g_{\rm r}$, and $h_{\rm r}$ about the CD as
\begin{equation}
f_{\rm r} = F\eta, \quad g_{\rm r} = G\eta^{\alpha}, \quad h_{\rm r} = H_0 + H_1\eta^{\beta}
\end{equation}
with $F$, $G$, $H_0$, $H_1$, $\alpha$, and $\beta$ constants; we demand that $H_0 > 0$ and $\beta > 0$ to ensure physical and self-consistent solutions. Inserting these relationships into the three self-similar Equations, \eqref{ssrcont} -- \eqref{ssrent}, yields the following four conditions that must be satisfied:
\begin{equation}
\begin{split}
&\left(2-m\right)-\left(1+\frac{m-n}{2}\right)\alpha+F\left(\alpha+1\right) = 0, \\
&\beta = 1+\alpha, \\
&\frac{H_1}{G}\beta = \frac{\kappa_{\rm c}}{\kappa_{\rm r}}\left(\frac{m-n}{2}\right)\left(1+\frac{m-n}{2}\right), \\
&m\gamma-n-\gamma\alpha\left(F-1-\frac{m-n}{2}\right) = 0. \label{asymptotes}
\end{split}
\end{equation}
The third of these is independent of the remaining three and is not needed to address the self-consistency of the solutions, but serves to demonstrate that the pressure increases (decreases) near the CD if $m > n$ ($m < n$), which is apparent from Figure \ref{fig:fr_gr_hr}. Combining the remaining three relations yields
\begin{equation}
\alpha = \frac{m\gamma-n}{n+\gamma\left(\frac{n-m}{2}-3\right)}, \quad \beta = \frac{\gamma\left(6-m-n\right)}{\gamma\left(m+6\right)-n\left(\gamma+2\right)}, \quad F = \frac{n-2\gamma}{\gamma}.
\end{equation}

\begin{figure}[htbp] 
   \centering
   \includegraphics[width=0.495\textwidth]{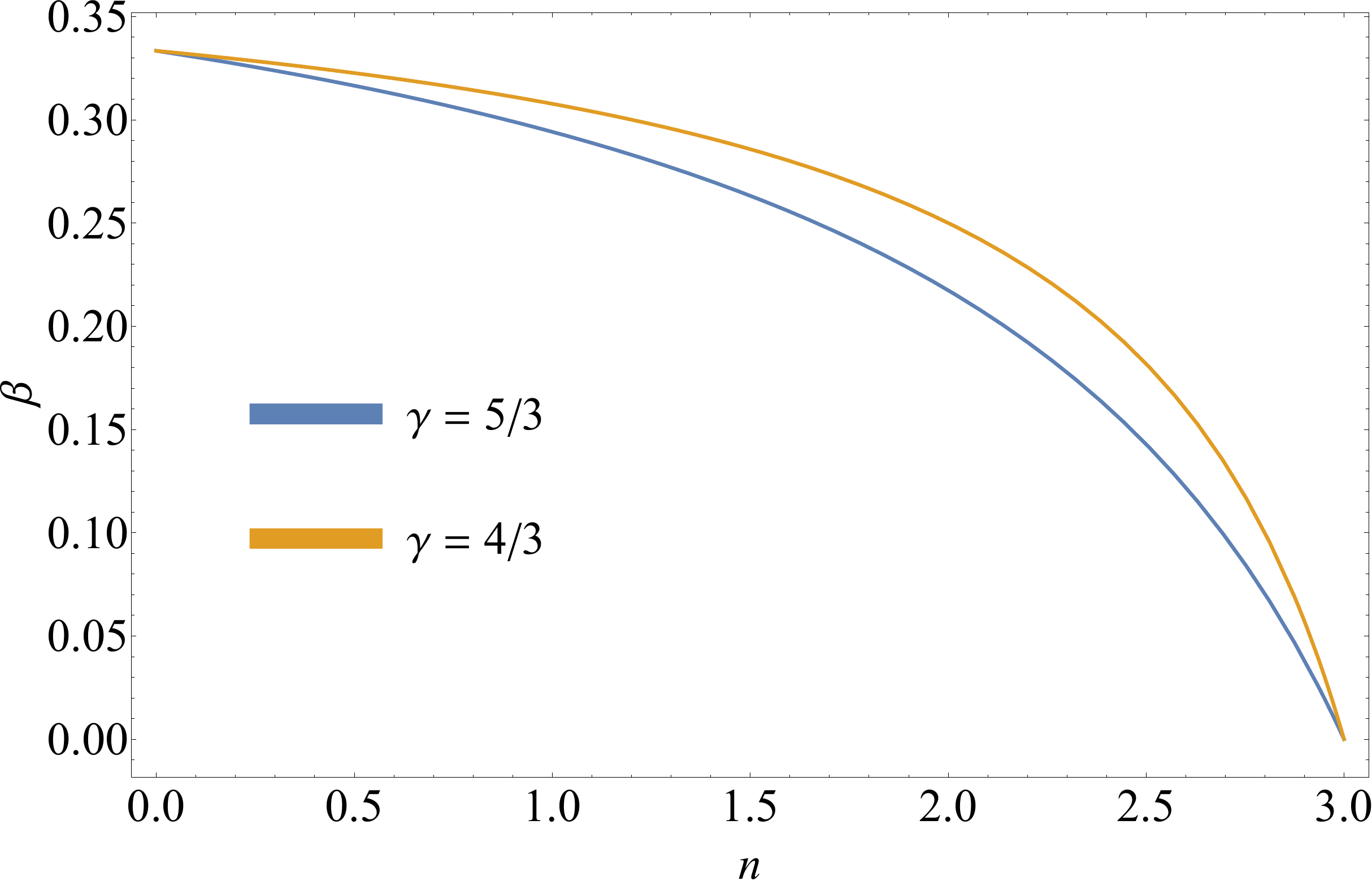} 
   \includegraphics[width=0.495\textwidth]{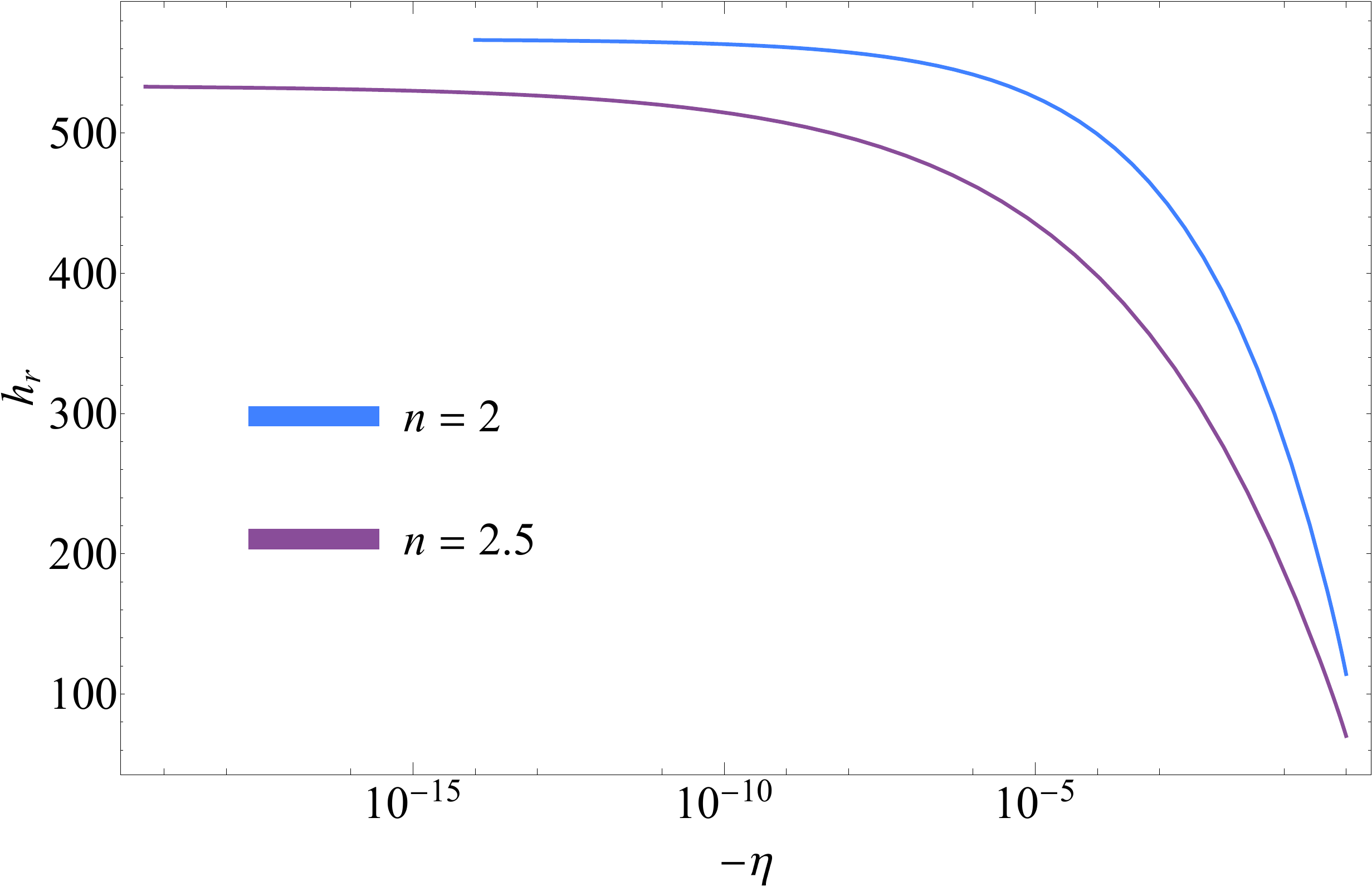} 
   \caption{Left: $\beta$ (the leading non-zero power-law index of the pressure) as a function of $n$ for $\gamma = 5/3$ (blue) and $\gamma = 4/3$ (orange/yellow) and $m = 3$. Since $\beta > 0$, the pressure converges near the CD, but the rate of convergence is extremely slow as $n \rightarrow 3$. Right: the self-similar pressure in the RS shell, $h_{\rm r}$, as a function of $-\eta$ on a logarithmic scale. The numerical accuracy needs to be extremely high, with the functions resolved at $|\eta| \le 10^{-11}$ for $n = 2$ and $|\eta| \le 10^{-16}$ for $n = 2.5$, to correctly determine the pressure at the CD and thus the eigenvalues controlling the expansion of the RS shell and the deceleration of the CD.}
   \label{fig:beta_of_n}
\end{figure}

The left panel of Figure \ref{fig:beta_of_n} shows $\beta$ for $\gamma = 5/3$ (blue) and $\gamma = 4/3$ (orange/yellow) for $m = 3$ as a function of $n$, which shows that the power-law index of the pressure near the CD is always positive, but tends toward zero as $n \rightarrow 3$. This shows that the pressure does converge to a finite value at the CD, but the very weak dependence on $\eta$, coupled to the fact that $H_1$ diverges in the same limit (see the third of Equation \eqref{asymptotes} and the fact that $H_1 \propto 1/\beta$), implies that extremely high numerical accuracy is required to correctly deduce that value. To demonstrate this directly, the right panel shows the self-similar pressure $h_{\rm r}$ throughout the RS shell, with $-\eta$ on the horizontal axis on a logarithmic scale. The pressure only converges to a constant value at extremely small values of $-\eta$: $|\eta| \lesssim 10^{-11}$ for $n = 2$ and $|\eta| \lesssim 10^{-16}$ for $n = 2.5$.

\bibliographystyle{aasjournal}

\end{document}